\documentclass[aps,prd,superscriptaddress,twocolumn,nofootinbib,floatfix]{revtex4-2}
\usepackage[normalem]{ulem}

\usepackage{natbib}
\usepackage{graphicx}
\usepackage{amsfonts}
\usepackage{amsmath,amssymb} 
\usepackage{bm}
\usepackage{xcolor}
\usepackage{hyperref}
\usepackage[caption=false]{subfig}
\usepackage{tabularx}
\usepackage{mathtools}
\usepackage{xfrac} 
\usepackage{orcidlink}
\usepackage{makecell} 

\newcommand{\En}{\mathcal{E}}

\newcommand{\Q}{\mathcal{Q}}

\renewcommand{\sp}{\epsilon} 
\newcommand{\mr}{\epsilon} 

\newcommand{\mycomment}[1]{} 

\newcommand{\avg}[1]{\left\langle #1 \right\rangle}

\newcommand{\osc}[1]{\Breve{#1}}
\newcommand{\nit}[1]{\Tilde{#1}}
\newcommand{\PD}[2]{\frac{\partial #1}{\partial #2}}
\newcommand{\HOT}[1]{\mathcal{O}( \sp ^{ #1 } )} 

\newcommand{\PaperI}{Paper~I}

\definecolor{aldo}{RGB}{210, 190, 100}

\renewcommand{\vec}[1]{\mathbf{#1}}

\begin{document}


\title{Efficient Eccentric and Quasicircular Effective-One-Body Waveforms via Near-Identity Averaging Transformations}


\newcommand{\NBIA}{Niels Bohr International Academy, Niels Bohr Institute, Blegdamsvej 17, 2100 Copenhagen, Denmark}
\newcommand{\AEI}{Max Planck Institute for Gravitational Physics (Albert Einstein Institute), Am M\"uhlenberg 1, Potsdam 14476, Germany}

\newcommand{\COG}{Center of Gravity, Niels Bohr Institute, Blegdamsvej 17, 2100 Copenhagen, Denmark}
\newcommand{\UofM}{Department of Physics, University of Maryland, College Park, MD 20742, USA}

\author{Philip Lynch \orcidlink{https://orcid.org/0000-0003-4070-7150}}
\email[Email: ]{philip.lynch@aei.mpg.de}
\affiliation{\AEI}

\author{Alessandra Buonanno \orcidlink{https://orcid.org/0000-0002-5433-1409}}
\affiliation{\AEI}
\affiliation{\UofM}
\author{Aldo Gamboa \orcidlink{https://orcid.org/0000-0001-8391-5596}} 
\affiliation{\AEI}
\author{Maarten van de Meent \orcidlink{https://orcid.org/0000-0002-0242-2464}}
\affiliation{\COG}
\affiliation{\AEI}

\date{\today}

\begin{abstract}
Next-generation gravitational-wave detectors, such as LISA, the Einstein Telescope, and Cosmic Explorer, will require accurate and efficient models of long-lived, eccentric black-hole binary signals. In previous work, we reduced the cost of evolving nonspinning eccentric effective-one-body dynamics by recasting the equations of motion in terms of osculating orbital elements and applying near-identity averaging transformations to remove the fast orbital-timescale structure during the inspiral. However, the parametrization in terms of the eccentricity and relativistic anomaly becomes singular in the circular-orbit limit, excluding quasicircular binaries and the late stages of eccentric inspirals. Moreover, the cost of generating the waveform from the averaged dynamics limited the overall speed-up to a factor of $1.5-8$. In this work, we replace the eccentricity and relativistic anomaly by the Cartesian components of the eccentricity vector, in which the averaged equations are regular for all eccentricities $0 \leq e < 1$. We also decompose the waveform modes into radial harmonics whose coefficients are tabulated as functions of the averaged variables, so that the waveform is assembled directly on the output time grid at a fixed cost per sample. We validate the method against the full effective-one-body evolution across the parameter space of binaries observable by ground-based detectors, now including circular orbits. At second postadiabatic order, the waveforms reach median mismatches of $\sim 10^{-8}$ with a maximum of $1.1 \times 10^{-5}$, while achieving speed-ups of $3-100 \times$ on a single core or up to $300 \times$ on eight cores against the serial EOB waveform. Given the speed-ups demonstrated here, we plan to apply these techniques to state-of-the-art SEOBNR models in future work.
\end{abstract}


\maketitle


\section{Introduction \label{section:Intro}}
In the decade since the first observation of gravitational waves from a merging binary black hole (BBH) \cite{LIGOScientific:2016aoc}, the LIGO-Virgo-KAGRA collaboration has reported more than 390 gravitational wave (GW) signals from coalescing stellar mass black holes and neutron stars \cite{LIGOScientific:2018mvr,LIGOScientific:2020ibl,KAGRA:2021vkt,LIGOScientific:2025slb,LIGOScientific:2026wfs}. These observations rest on waveform models used to find the signals in the data, to infer the properties of their sources, and to test for deviations from general relativity. The next generation of ground-based detectors, the Einstein Telescope \cite{ET:2025xjr} and Cosmic Explorer \cite{Reitze:2019iox}, and space-based detectors such as the Laser Interferometer Space Antenna (LISA) \cite{LISA:2024hlh,LISAConsortiumWaveformWorkingGroup:2023arg}, TianQin \cite{TianQin}, and Taiji \cite{Taiji}, will provide even greater sensitivity, probe new frequency ranges and uncover entirely new sources of gravitational waves. However, effects that are currently buried in the noise are expected to become sources of systematic error if not sufficiently accounted for \cite{Littenberg:2012uj, Purrer:2019jcp, Owen:2023mid,Dhani:2024jja,Kolitsidou:2024vub,MacUilliam:2024oif}.

Orbital eccentricity is one such effect. Although it has not yet been established beyond doubt in any individual event, hints of eccentricity have been reported in a growing number of detected signals \cite{Gayathri:2020coq,Romero-Shaw:2021ual,Romero-Shaw:2022xko,Iglesias:2022xfc,Ramos-Buades:2023yhy,Gupte:2024jfe,Morras:2025xfu,Planas:2025jny,Xu:2025ajj,Gupte:2026whi,Zeeshan:2026pga,Pompili:2026yxq,Gamboa:2026ybr}. Gravitational radiation circularizes a binary as it inspirals \cite{Peters:1963ux,Peters:1964zz,Hinder:2007qu,Sperhake:2007gu}, but dynamical formation in dense environments such as globular clusters and active galactic nuclei can leave binaries with eccentricities that are still measurable when they enter the sensitivity band of present detectors \cite{Downing_2010,Antonini:2012ad,OLeary:2008myb,Petrovich:2017otm,Gondan:2017wzd,Takatsy:2018euo}, and even more so for next-generation detectors, which are sensitive to the early inspiral where eccentricity leaves its clearest imprint \cite{Saini:2023wdk, Nishizawa:2016jji, Nishizawa:2016eza, Garg:2023lfg}. At the same time, the majority of detected binaries are consistent with circular orbits, so that an eccentric waveform model must remain accurate and efficient all the way down to, and including, the circular limit.

Consequently, orbital eccentricity has been incorporated into most of the waveform-modeling approaches in use today \cite{Huerta:2016rwp,Huerta:2017kez,Hinder:2017sxy,Chattaraj:2022tay,Manna:2024ycx,Paul:2024ujx,Islam:2022laz,Nee:2025nmh,Maurya:2025shc,Morras:2025nlp,Planas:2025feq,Ramos-Buades:2026kbq,Morras:2026fho,Islam:2026blk}. Here we work within the effective-one-body (EOB) framework \cite{Buonanno:1998gg,Buonanno:2000ef,Damour:2000we,Buonanno:2005xu,Barausse:2009xi}, which maps the two-body problem onto a test mass moving in a deformed black hole spacetime and resums information from post-Newtonian (PN) \cite{Nagar:2011fx,Bini:2012ji,Balmelli:2013zna,Damour:2015isa,Khalil:2020mmr,Khalil:2021txt,Gamboa:2024imd}, gravitational self-force (GSF) \cite{Antonelli:2019fmq,vandeMeent:2023ols,Leather:2025nhu}, and post-Minkowskian (PM) calculations \cite{Damour:2016gwp,Antonelli:2019ytb,Khalil:2022ylj,Buonanno:2024byg,Buonanno:2024vkx,Damour:2025uka} into this effective description. Calibration to numerical relativity (NR) simulations then improves the accuracy of the late inspiral and the merger-ringdown \cite{Pan:2009wj,Buonanno:2009qa,Bohe:2016gbl}. The resulting models are among the most accurate semi-analytic models available and have seen widespread adoption in GW data analysis \cite{Nagar:2020pcj,Gamba:2021ydi,Pompili:2023tna,Khalil:2023kep,Ramos-Buades:2023ehm,Mihaylov:2023bkc}. Recently, considerable effort has gone into extending these models to eccentric orbits \cite{Nagar:2021gss,Nagar:2024dzj,Nagar:2024oyk,Liu:2021pkr,Ramos-Buades:2021adz,Gamboa:2024hli,Gamboa:2026jht} and to precessing spins \cite{Liu:2023ldr,Gamba:2024cvy,Albanesi:2025txj,Gamboa:2026ybr}.

The price of this accuracy is that the inspiral has to be computed by numerically integrating the EOB Hamiltonian equations of motion. The notion of a ``postadiabatic expansion'' around quasicircular orbits in EOB was first used to derive the transition to plunge \cite{Buonanno:2000ef,Buonanno:2005xu} and then to set more accurate initial conditions for NR comparisons \cite{Damour:2007yf,Damour:2012ky}. Ref.~\cite{Nagar:2018gnk} exploited this expansion to transform the equations of motion into algebraic relations that are solved iteratively on a radial grid, removing the need to integrate the dynamics and accelerating production quasicircular EOB waveform models \cite{Rettegno:2019tzh,Riemenschneider:2021ppj,Mihaylov:2021bpf}. No such shortcut exists for eccentric binaries, whose dynamics must be integrated numerically to resolve every orbital cycle. As such, the expense grows with the number of orbits in band, i.e.\ it is largest for the low-mass and large-mass-ratio systems that spend the longest in the detector.

In Ref.~\cite{Lynch:2026ibo} (henceforth \PaperI) we addressed this problem by adapting techniques that have proven highly successful in the modeling of eccentric extreme-mass-ratio inspirals (EMRIs) \cite{Chua:2020stf,Katz:2021yft,Speri:2023jte,Chapman-Bird:2025xtd}. We recast the eccentric EOB dynamics in terms of osculating orbital elements (OOE) \cite{Pound:2007th,Gair:2010iv} (also known as the method of variation of constants \cite{taff1985celestial,beutler2005volume1}) using a quasi-Keplerian parametrization à la Ref.~\cite{Hinderer:2017jcs}, and then applied near-identity (averaging) transformations (NITs) \cite{Kevorkian1987} to recast the short-timescale oscillatory structure of the inspiral dynamics as higher-order secular information \cite{vandeMeent:2018cgn,McCart:2021upc,Lynch:2021ogr,Pound2022,Lynch:2022zov,Lynch:2023gpu,Drummond:2023wqc,Lynch:2024ohd,Drummond:2026haw}. The result is a set of averaged equations of motion, carried beyond the next-to-leading order needed for EMRIs, through second postadiabatic (2PA) order in the small parameter $\epsilon$ that characterizes the separation between the orbital and radiation-reaction timescales. These equations can be solved on the radiation-reaction timescale rather than the orbital timescale, and the solution is then mapped back to the full EOB dynamics for the late inspiral. Tested on an inspiral-only, nonspinning EOB model, the 2PA averaged dynamics reproduced the full EOB waveforms to mismatches $\lesssim 10^{-4}$ while speeding up the dynamics by one to two orders of magnitude.

Two issues remained to be addressed. The first is a limitation of the implementation of \PaperI, while the second was deliberately deferred to this work.
First, the quasi-Keplerian parametrization in terms of the eccentricity $e$ and the relativistic anomaly $\zeta$ \cite{Darwin:1961zz} is singular in the circular-orbit limit. The argument of periapsis $w$, and with it $\zeta = \chi - w$, where $ \chi $ is a radial phase parameter, is ill-defined as $e \rightarrow 0$. As a result, the equations of motion contain terms that diverge as $1/e$, and the oscillatory part of the osculating eccentricity does not vanish as the orbit circularizes, so that the osculating eccentricity can no longer be written as a small oscillation about the averaged one. We were therefore forced to hand over to the full EOB dynamics whenever the averaged eccentricity dropped below $e = 0.035$, excluding precisely the quasicircular systems that make up the bulk of the detected population and whose long inspirals would benefit most from the method. Second, once the dynamics are fast, the generation of the waveform itself becomes the bottleneck. In \PaperI\ we retained the conventional approach of evaluating the mode expressions at every point of a dense sampling grid, which limited the overall speed-up to a factor of $1.5$--$8$ despite a speed-up of the dynamics of up to two orders of magnitude, and left the development of a faster waveform-generation scheme to this paper.

In this work, we address both issues. First, we replace the eccentricity and relativistic anomaly by the Cartesian components of the eccentricity vector $\alpha = e \sin w$ and $\beta = e \cos w$, which are regular through the circular-orbit limit, as announced in \PaperI. These are the two in-plane equinoctial elements of celestial mechanics \cite{taff1985celestial}, which have also proved useful for the modeling of compact binaries \cite{Pound:2007th,Will:2019lfe,Tucker:2021mvo}. We rederive the osculating-element equations of motion of the nonspinning EOB dynamics and the NITs through 2PA order in these variables. Moreover, we show that the rotational covariance of the equations under a shift of the periastron reduces the resulting four-dimensional parameter space back to the three-dimensional grid of \PaperI, so that its interpolated Fourier coefficients can be reused. 

Second, we introduce a new waveform generation method inspired by the multi-voice decomposition common in EMRI waveform modeling \cite{Hughes:2021exa,Katz:2021yft,Chua:2020stf,Speri:2023jte,Chapman-Bird:2025xtd}. The waveform modes are decomposed into radial harmonics whose coefficients are tabulated across the parameter space directly as functions of the averaged variables. The waveform is then assembled on the output time grid by inverting the time transformation sample by sample, with no evaluation of the mode expressions along the inspiral. Remarkably, this process also lowers the $\sim 10^{-6}$--$10^{-7}$ mismatch floor reached with the method of \PaperI.

Finally, we validate the method against the full EOB evolution of the same inspiral-only, nonspinning model over the parameter space of \PaperI, now extended to circular orbits. The 2PA waveforms reproduce the EOB waveforms with median mismatches of $\sim 10^{-8}$ and maximum mismatches of $\sim 10^{-5}$ for total masses between $10$ and $100 M_\odot$, with no loss of accuracy in the circular limit. The full waveform is generated $3$--$100$ times faster on a single core and up to $300$ times faster when it is distributed across $8$ cores and compared with the serial EOB waveform. The speed-up is greatest for low-mass, large-mass-ratio systems, which are the most computationally expensive to evolve and the most likely to benefit from further parallelization and GPU acceleration. The inspiral-only, nonspinning model used here is a deliberately simplified version of the \texttt{SEOBNR} models, chosen as a proof of principle. The method itself places no restriction on the underlying EOB dynamics, and we expect it to extend readily to the state-of-the-art spinning \texttt{SEOBNR} models, which we leave to future work.

The remainder of this paper is organized as follows. In Sec.~\ref{section:EccentricDynamics}, we review the nonspinning eccentric EOB dynamics and their osculating-element formulation from \PaperI, and introduce the $(\alpha,\beta)$ parametrization in which we rewrite the osculating-element equations. In Sec.~\ref{section:NITs_QC}, we derive the near-identity averaged equations of motion through second postadiabatic order in the new variables and show how the parameter space is reduced. In Sec.~\ref{section:Waveforms}, we describe the Fourier decomposition of the waveform modes and the assembly of the waveform on the time grid, and in Sec.~\ref{section:Implementation} we summarize the numerical implementation. We present accuracy and runtime comparisons in Sec.~\ref{section:Results}, and conclude in Sec.~\ref{section:Discussion} with a discussion of limitations and future extensions. In this work, we use geometric units where $G = c = 1$.

\section{Nonspinning eccentric dynamics} \label{section:EccentricDynamics}
In this section, we review the nonspinning eccentric EOB dynamics and their reformulation in terms of osculating orbital elements from \PaperI, and then introduce the equinoctial parametrization that we use in the remainder of this work. We refer the reader to \PaperI\ for the full derivations and explicit expressions in the original variables.

\subsection{Nonspinning effective-one-body dynamics} \label{section:EOB_Dynamics}
We consider a two-body system consisting of a primary of mass $m_1$ and a secondary of mass $m_2$, with total mass $M = m_1 + m_2$, reduced mass $\mu = m_1 m_2 / M$, and symmetric mass ratio $\nu = \mu / M$. 
In the EOB framework, the binary is modeled as a test mass moving in a deformed black hole spacetime where the deformation parameter is $\nu$. The metric of this deformed spacetime in EOB coordinates $(t,r,\phi)$ is
\begin{equation}
	ds^2 = -A(\nu;r) dt^2 + \frac{dr^2}{A(\nu;r) \bar{D}(\nu;r)} + r^2 d\phi^2,
\end{equation}
which reduces to the Schwarzschild metric in the test mass limit $\nu \rightarrow 0$. As in \PaperI, we adopt the potentials $A$ and $\bar{D}$ of \texttt{SEOBNRv5} \cite{Pompili:2023tna,Khalil:2023kep}.

The effective Hamiltonian is given by
\begin{equation} \label{eq:Heff}
	H_{\text{eff}}^2 = p_{r_*}^2 + A(\nu;r)\left[1 + \frac{p_\phi^{2}}{r^2} + Q(\nu;r,p_{r_*})\right],
\end{equation}
where $p_{r_*} = \xi(r) p_r$ is the tortoise radial momentum with $\xi(r) = A \sqrt{\bar{D}}$, and $Q$ is a PN series in $p_{r_*}$ beginning at $\mathcal{O}(p_{r_*}^4)$ that is known to $\mathcal{O}(p_{r_*}^8)$ \cite{Pompili:2023tna}. All momenta are normalized by $\mu$. The effective Hamiltonian is related to the EOB Hamiltonian via the energy map
\begin{equation}
	H_\text{EOB} = M \sqrt{1 + 2 \nu (H_\text{eff} -1)},
\end{equation}
from which the equations of motion follow via Hamilton's equations:
\begin{align} \label{eq:EOB_EOM}
	\begin{split}
	\frac{d r}{dt} &= \xi(r) \frac{\partial H_{\text{EOB} }}{\partial p_{r_*}}, \quad \frac{d \phi}{dt} = \frac{\partial H_{\text{EOB}}}{\partial p_\phi} = \Omega_\phi,
	\end{split}\\
\begin{split}
	\frac{d p_{r_*}}{dt} &= - \xi(r) \frac{\partial H_{\text{EOB}}}{\partial r} + \xi(r) F_r, \quad \frac{d p_\phi}{dt} = F_\phi.
	\end{split}
\end{align}
Here $F_r$ and $F_\phi$ are the radiation-reaction force components, for which we use the nonspinning limit of the 3PN expressions of Ref.~\cite{Gamboa:2024imd} (given explicitly in Appendix~C of \PaperI).

\subsection{Osculating orbital elements for eccentric orbits} \label{section:OOE_Summary}
In \PaperI, we recast the EOB equations of motion into evolution equations for a set of osculating orbital elements \cite{Pound:2007th,Gair:2010iv} using the quasi-Keplerian parametrization \cite{Hinderer:2017jcs}
\begin{equation} \label{eq:r_of_p_e_zeta}
	r = \frac{p M}{1 + e \cos \zeta},
\end{equation}
where $p$ is the semilatus rectum, $e$ is the eccentricity, and $\zeta = \chi - w$ is the relativistic anomaly, with $\chi$ the radial phase parameter and $w$ the argument of periapsis. The semilatus rectum and eccentricity are related to the radial turning points $r_1$ and $r_2$ of the conservative orbit via
\begin{equation}
	p = \frac{2 r_1 r_2}{M(r_1 + r_2)}, \quad 
	e = \frac{r_1 - r_2}{r_1 + r_2}.
\end{equation}
The effective energy $H_\text{eff}(\nu;p,e)$ and angular momentum $p_\phi(\nu;p,e)$ of the tangent conservative orbit are functions of $p$ and $e$ only, while the radial momentum $p_{r_*}(\nu;p,e,\zeta)$ is obtained by perturbatively inverting Eq.~\eqref{eq:Heff} (see Appendix~A of \PaperI). 

The orbital elements $I^A = (p,e,w)$ are constant in the absence of radiation reaction. The method of osculating orbital elements maps the inspiraling worldline at each instant onto the tangent conservative orbit with the same orbital elements. 
From this, one obtains the two osculating conditions
\begin{subequations} \label{eq:osculating_conditions}
	\begin{align}
		\PD{x^\alpha}{I^A} \frac{d I^A}{d t} &= 0, \label{eq:osculating_position} \\
		\PD{p_\alpha}{I^A} \frac{d I^A}{d t} &= F_\alpha, \label{eq:osculating_momentum}
	\end{align}
\end{subequations}
where $x^\alpha = (r,\phi)$ and $p_\alpha = (p_r, p_\phi)$ are the coordinates and momenta of the tangent conservative orbit, expressed as functions of the orbital elements and the phase, and $F_\alpha = (F_r, F_\phi)$ are the components of the radiation-reaction force. The first condition states that the variation of the orbital elements leaves the position unchanged, and the second that the momenta change only through the radiation-reaction force. Solving these conditions for $dI^A/dt$ gives the evolution equations of the orbital elements, which are given explicitly in Sec.~III of \PaperI.

In \PaperI, we parametrized the system by the radial phase $\chi$ rather than the time $t$, and absorbed the evolution of $w$ into that of $\zeta = \chi - w$. The resulting equations of motion take the form
\begin{subequations} \label{eq:OOE_EoM_Eccentric_Param}
	\begin{align}
		\begin{split}
			& \dot{P_j} = \epsilon F_j (\nu;\vec{P}, \zeta) ,
		\end{split}\\
		\begin{split}
			& \dot{\zeta} = 1 + \epsilon f_{w}(\nu; \vec{P}, \zeta),
		\end{split}\\
		\begin{split}
			& \dot{S}_k = s_k(\nu;\vec{P}, \zeta),
		\end{split}
	\end{align}
\end{subequations}
where $P_j = (p,e)$ are the orbital elements, $S_k = (t,\phi)$ are the ``extrinsic'' quantities, an overdot denotes the rate of change with respect to $\chi$, and $\epsilon$ is a bookkeeping parameter that marks terms which depend on radiation reaction. Explicitly, $F_p = (dp/dt)(dt/d\chi)$, $F_e = (de/dt)(dt/d\chi)$, $\epsilon f_w = - (dw/dt)(dt/d\chi)$, $s_t = f_t = dt/d\chi$, and $s_\phi = f_\phi = (d\phi/dt)(dt/d\chi)$.

The forcing functions $F_j$, $f_w$, and $s_k$ are periodic in $\zeta$, with unphysical $0/0$ singularities at the radial turning points that we regularize by Fourier decomposition (Sec.~\ref{section:FourierDecomposition_QC}).

This parametrization is ill-suited to small eccentricities. The argument of periapsis $w$ is ill-defined for a circular orbit, so that $f_w$ diverges as $1/e$, and the evolution equations for $p$ and $e$ involve the inverse of the Jacobian of the transformation from $(H_\text{eff},p_\phi)$ to $(p,e)$, which is also singular as $e \rightarrow 0$. Moreover, as we show in Appendix~\ref{appendix:OOE_EoM_QC}, the osculating eccentricity keeps oscillating with a finite amplitude as the orbit circularizes. In \PaperI\ this forced a handover from the averaged dynamics back to the full EOB dynamics whenever the eccentricity dropped below $e = 0.035$, which is the primary limitation we address here.

\subsection{Equinoctial parametrization} \label{section:OOE_QC}
To remove this limitation, we replace the eccentricity and argument of periapsis by the two in-plane equinoctial elements of celestial mechanics \cite{taff1985celestial}, which we refer to as the $(\alpha,\beta)$ parametrization throughout. We define the new orbital elements
\begin{equation} \label{eq:alpha_beta_def}
	\alpha = e \sin w, \qquad \beta = e \cos w,
\end{equation}
which are the Cartesian components of the eccentricity vector in the orbital plane. The original orbital elements are recovered via
\begin{equation} \label{eq:e_w_of_alpha_beta}
	e = \sqrt{\alpha^2 + \beta^2}, \qquad \tan w = \frac{\alpha}{\beta}.
\end{equation}
Substituting Eq.~\eqref{eq:alpha_beta_def} into Eq.~\eqref{eq:r_of_p_e_zeta} with $\zeta = \chi - w$, the radial coordinate becomes
\begin{equation} \label{eq:r_of_p_alpha_beta}
	r = \frac{p M}{1 + \alpha \sin \chi + \beta \cos \chi}.
\end{equation}

The key advantage of this parametrization is that $e = 0$ is a regular point of the $(\alpha,\beta)$ plane rather than a coordinate singularity. The argument of periapsis $w$, which is ill-defined for a circular orbit, no longer appears as an independent variable. Furthermore, since the effective energy $H_\text{eff}(\nu;p,e)$ and angular momentum $p_\phi(\nu;p,e)$ of the tangent conservative orbit depend only on $p$ and $e$, they are unchanged from \PaperI\ upon substituting Eq.~\eqref{eq:e_w_of_alpha_beta}. Likewise, the radial momentum is given by $p_{r_*}(\nu;p,e,\zeta)$ from Appendix~A of \PaperI\ with $\zeta = \chi - w$.

In this parametrization, the radial phase $\chi$ is no longer decomposed into a relativistic anomaly and an argument of periapsis. Instead, $\chi$ is the independent variable of the system, and remains well defined in the limit $e \rightarrow 0$, where it reduces to the phase of the radial epicyclic oscillation about the circular orbit. We examine the quasicircular limit of the osculating orbital element equations in Appendix~\ref{appendix:OOE_EoM_QC}.

The evolution equations for $\alpha$ and $\beta$ follow from those of $e$ and $w$ via the chain rule. Using the equations of motion of \PaperI\ in the form of Eq.~\eqref{eq:OOE_EoM_Eccentric_Param}, i.e., $de/d\chi = \epsilon F_e$ and $dw/d\chi = -\epsilon f_w$, we find
\begin{subequations} \label{eq:F_alpha_beta}
	\begin{align}
	\begin{split}
		\frac{d \alpha}{d \chi} &= \mr \left( F_e \sin w - e f_{w} \cos w \right) \equiv \mr F_\alpha,
	\end{split}\\
	\begin{split}
		\frac{d \beta}{d \chi} &= \mr \left( F_e \cos w + e f_{w} \sin w \right) \equiv \mr F_\beta.
	\end{split}
	\end{align}
\end{subequations}
Note that while $f_w$ diverges as $1/e$ in the circular-orbit limit, the combination $e f_w$ that appears in the above expressions remains finite, so that $F_\alpha$ and $F_\beta$ are well behaved as $e \rightarrow 0$.

Overall, we obtain equations of motion of the form
\begin{subequations} \label{eq:OOE_EoM_QC_Explicit}
\begin{align}
	\begin{split}
		\frac{d p}{d \chi} &=  \epsilon F_p(\nu;p,\alpha, \beta,\chi), 
	\end{split} \\
	\begin{split}
		\frac{d \alpha}{d \chi} &= \epsilon F_\alpha(\nu;p,\alpha, \beta,\chi), 
	\end{split}\\
	\begin{split}
		\frac{d \beta}{d \chi} &=  \epsilon F_{\beta} (\nu;p,\alpha, \beta,\chi), 
	\end{split} \\
	\begin{split}
		\frac{d t}{d \chi} & = f_t  (\nu;p,\alpha, \beta,\chi),
	\end{split}\\
	\begin{split}
		\frac{d \phi}{d \chi} &=  f_\phi   (\nu;p,\alpha, \beta,\chi),
	\end{split}
\end{align}
\end{subequations}
where the forcing functions of \PaperI\ have been re-expressed in terms of $(p,\alpha,\beta,\chi)$ using Eqs.~\eqref{eq:e_w_of_alpha_beta}. We summarize these schematically as
\begin{subequations}\label{eq:OOE_EoM_QC}
	\begin{align}
		\begin{split}
			\dot{P_j} &= \epsilon F_j (\nu;\vec{P}, \chi) ,
		\end{split}\\
		\begin{split}
			\dot{S}_k &= s_k(\nu;\vec{P}, \chi),
		\end{split}
	\end{align}
\end{subequations}
where now the orbital elements are $P_j = (p,\alpha,\beta)$ and, as before, $S_k = (t,\phi)$. In contrast to Eq.~\eqref{eq:OOE_EoM_Eccentric_Param}, there is no longer a separately evolved orbital phase; instead, we have an additional orbital element, and the role of the orbital phase is played directly by the independent variable $\chi$.

We stress that this parametrization is not restricted to small eccentricities. Eqs.~\eqref{eq:OOE_EoM_QC} are exactly equivalent to Eqs.~\eqref{eq:OOE_EoM_Eccentric_Param} for all $0 < e < 1$ and, unlike the latter, remain regular at $e = 0$, so we use them for the entire inspiral at all eccentricities. The price is that the intrinsic parameter space now consists of three variables $(p,\alpha,\beta)$ rather than two $(p,e)$, which would make sampling and interpolating across the parameter space more expensive. As we show in Sec.~\ref{section:ParameterSpaceReduction}, the dependence on the additional variable is fixed analytically by a rotational symmetry, so that the interpolation tables remain three-dimensional.

\subsection{Regularization of forcing terms via Fourier decomposition} \label{section:FourierDecomposition_QC}
As in \PaperI, the forcing functions $F_j$ and $s_k$ are periodic functions of $\chi$ with period $2\pi$, but possess unphysical numerical singularities at the radial turning points, now located at $\chi = w + n \pi$ for $n \in \mathbb{Z}$. We regularize them by decomposing into Fourier series with respect to $\chi$:
\begin{equation} \label{eq:Fourier_QC}
	A(\vec{P},\chi) = \sum_{\kappa} A_{\kappa}(\vec{P}) e^{i  \kappa \chi},
\end{equation}
where $A$ represents any of the functions $F_j$ or $s_k$. Based on this, we split each function into an averaged piece
\begin{equation}\label{eq:average_QC}
	\avg{A} (\vec{P}) = A_{0}(\vec{P}) = \frac{1}{2\pi} \int_0^{2\pi} A(\vec{P},\chi) d \chi,
\end{equation}
and an oscillating piece
\begin{equation}\label{eq:oscillating_QC}
	\osc{A}(\vec{P},\chi) = A(\vec{P},\chi) - \avg{A}(\vec{P})  = \sum_{\kappa \neq 0 } A_{\kappa}(\vec{P}) e^{i \kappa  \chi}. 
\end{equation}
The Fourier coefficients are computed via a fast Fourier transform, sampling at equal steps in $\chi$ with a constant offset to avoid the turning points, and the slowly decaying coefficients of $f_t$ and $f_\phi$ are computed via their reciprocals as described in Appendix~B of \PaperI. Since all coefficients are computed on the slice $w = 0$ (Sec.~\ref{section:ParameterSpaceReduction}), where the turning points sit at $\chi = n\pi$ as in \PaperI, the same fixed offset from periastron can be used. The interpolation grid is described in Sec.~\ref{section:InterpolationGrid}.

\section{Near-identity averaging transformations in the equinoctial parametrization} \label{section:NITs_QC}
Near-identity (averaging) transformations (NITs) \cite{Kevorkian1987} are small transformations of the equations of motion that average out the short timescale physics while retaining the long term evolution. First applied in gravitational-wave modeling to extreme-mass-ratio inspirals \cite{vandeMeent:2018cgn,McCart:2021upc,Lynch:2021ogr,Pound2022,Lynch:2022zov,Lynch:2023gpu,Drummond:2023wqc,Lynch:2024ohd,Drummond:2026haw}, in \PaperI\ we showed that they are also well suited to accelerating eccentric EOB inspiral dynamics. We now apply them to the osculating orbital element equations in the $(\alpha,\beta)$ parametrization, Eqs.~\eqref{eq:OOE_EoM_QC}. The procedure closely parallels that of \PaperI, with one important simplification. Since there is no longer a separately evolved orbital phase, no transformation of the phase variable is required.

We define NIT variables $\nit{P}_j$ and $\nit{S}_k$, related to the OOE variables $P_j$ and $S_k$ via
\begin{subequations}\label{eq:transformation_QC}
\begin{align}
	\begin{split}\label{eq:transformation_QC_P}
		\nit{P}_j &= P_j + \sp Y_j^{(1)}(\vec{P}, \chi) + \sp^2 Y_j^{(2)}(\vec{P}, \chi) +  \HOT{3},
	\end{split}\\
	\begin{split}\label{eq:transformation_QC_S}
		\nit{S}_k &= S_k + Z_k^{(0)}(\vec{P}, \chi) +\sp Z_k^{(1)}(\vec{P},  \chi) + \HOT{2}.
	\end{split}
\end{align}
\end{subequations}
Here, the transformation functions $Y_j^{(n)}$ and $Z_k^{(n)}$ are required to be smooth, periodic functions of $\chi$, and also depend on $\nu$, which we suppress. At leading order, Eqs.~\eqref{eq:transformation_QC} are identity transformations for $P_j$ but not for $S_k$, owing to the zeroth-order transformation term $Z_k^{(0)}$.
The inverse transformations are found by requiring that their composition with Eqs.~\eqref{eq:transformation_QC} recovers the identity transformation. Expanding order by order in $\epsilon$, this gives
\begin{subequations}\label{eq:inverse_transformation_QC}
	\begin{align}
	\begin{split}
		P_j &= \nit{P_j} - \epsilon Y_j^{(1)}(\vec{\nit{P}},\chi) - \epsilon^2 \nit{Y}_j^{(2)}(\vec{\nit{P}},\chi) + \HOT{3},
	\end{split}\\
	\begin{split}\label{eq:inverse_transformation_QC_S}
		S_k  &= \nit{S}_k -Z_k^{(0)}(\vec{\nit{P}},\chi) - \epsilon \nit{Z}_k^{(1)}(\vec{\nit{P}},\chi) + \HOT{2},
	\end{split}
	\end{align}
\end{subequations}
where 
\begin{subequations}\label{eq:inverse_transformation_definitions_QC}
	\begin{align}
	\begin{split}
		\nit{Y}_j^{(2)} &= Y_j^{(2)} - \PD{Y_j^{(1)}}{\nit{P_k}} Y_k^{(1)} ,
	\end{split}\\
	\begin{split}
		\nit{Z}_k^{(1)} &= Z_k^{(1)} - \PD{Z_k^{(0)}}{\nit{P_j}} Y_j^{(1)} .
	\end{split}
	\end{align}
\end{subequations}
Compared with the corresponding expressions in \PaperI, the terms involving the phase transformation $X^{(1)}$, which is singular in the circular limit, are absent.

To derive the averaged equations of motion, one takes the $\chi$ derivative of Eqs.~\eqref{eq:transformation_QC}, substitutes the OOE equations of motion \eqref{eq:OOE_EoM_QC}, converts from OOE variables to NIT variables using Eqs.~\eqref{eq:inverse_transformation_QC}, and expands order by order in $\mr$. At each order, one then fixes the oscillating pieces of the transformation functions $Y_j^{(n)}$ and $Z_k^{(n)}$ such that the resulting equations of motion for $\nit{P}_j$ and $\nit{S}_k$ take the form
\begin{subequations}\label{eq:transformed_EoM_QC}
	\begin{align}
	\begin{split}
		\dot{\nit{P}}_j &= \epsilon \nit{F}_j^{(0)}(\vec{\nit{P}}) + \epsilon^2 \nit{F}_j^{(1)}(\vec{\nit{P}}) + \epsilon^3 \nit{F}_j^{(2)}(\vec{\nit{P}}) +  \HOT{4},
	\end{split}\\
	\begin{split}
		\dot{\nit{S}}_k &= \nit{s}_k^{(0)}(\vec{\nit{P}}) + \sp \nit{s}_k^{(1)}(\vec{\nit{P}}) + \sp^2  \nit{s}_k^{(2)}(\vec{\nit{P}})  +  \HOT{3},
	\end{split}
	\end{align}
\end{subequations}
where the superscripts in parentheses indicate the postadiabatic (PA) order of the terms. Following \PaperI, we refer to the leading order of the NIT expansion as the adiabatic order (0PA), to the next-to-leading order as first postadiabatic order (1PA), and to the next-to-next-to-leading order as second postadiabatic order (2PA). We do not expand in the mass ratio, so this usage differs from that of gravitational self-force theory \cite{Hinderer:2008dm}. It also differs from the quasicircular postadiabatic approximation of Refs.~\cite{Nagar:2018gnk,Rettegno:2019tzh,Riemenschneider:2021ppj,Mihaylov:2021bpf} as we are not expanding about a circular orbit. Crucially, these equations of motion are now independent of $\chi$. As before, the orbit averaged pieces of the transformation functions are unconstrained by this procedure, and we set them to zero. The derivation is provided in an ancillary Mathematica file, and we summarize the results below.

\subsection{Adiabatic terms}
The adiabatic (0PA) averaged terms are given by
\begin{subequations} \label{eq:0PA_Averaged_Terms_QC}
	\begin{align}
	\begin{split}
		\nit{F}_j^{(0)} &= \avg{F_{j}},
	\end{split}\\
	\begin{split} \label{eq:0PA_f_t_f_phi_QC}
		\nit{s}_k^{(0)} &=  \avg{s_{k}},
	\end{split}
	\end{align}
\end{subequations}
where the average is now taken with respect to $\chi$ as in Eq.~\eqref{eq:average_QC}. In deriving these relations, we have constrained the oscillating pieces of the NIT transformation functions to be
\begin{subequations} \label{eq:NIT_Y1_Z0_QC}
	\begin{align}
	\begin{split}\label{eq:NIT_Y1_QC}
		\osc{Y}_j^{(1)} & = \sum_{\kappa \neq 0} \frac{i}{\kappa} F_{j,\kappa} e^{i \kappa \chi},
	\end{split} \\
	\begin{split}\label{eq:NIT_Z0_QC}
		\osc{Z}_k^{(0)}  & = \sum_{\kappa \neq 0} \frac{i}{\kappa} s_{k,\kappa}e^{i \kappa \chi}.
	\end{split}
	\end{align}
\end{subequations}
As in \PaperI, we refer to the model that uses only the 0PA equations of motion together with the leading-order transformations for the initial conditions as the ``0PA model.''

\subsection{First postadiabatic terms}
At first postadiabatic order (1PA), one obtains
\begin{subequations} \label{eq:1PA_Averaged_Terms_QC}
	\begin{align}
	\begin{split}
		\nit{F}_j^{(1)} & = \avg{\PD{\osc{Y}_j^{(1)}}{\nit{P}_k} \osc{F}_k },
	\end{split}\\
	\begin{split}
		\nit{s}_k^{(1)} & =  \avg{\PD{\osc{Z}_k^{(0)}}{\nit{P}_j} \osc{F}_j },
	\end{split}
	\end{align}
\end{subequations}
where we have constrained
\begin{subequations} \label{eq:NIT_Y2_Z1_QC}
\begin{align}
	\begin{split}\label{eq:NIT_Y2_QC}
		\osc{Y}_j^{(2)}  = & \sum_{\kappa \neq 0} \frac{ i e^{ i \kappa \chi}}{\kappa}  \Biggl[ -  \PD{\nit{F}_j^{(0)}}{\nit{P}_k} \osc{Y}_{k, \kappa}^{(1)} + 
		  \sum_{\kappa' \neq 0}  \PD{\osc{Y}_{j, \kappa'}^{(1)}}{\nit{P}_k} \osc{F}_{k,\kappa - \kappa'} \Biggr],
	\end{split} \\
	\begin{split}\label{eq:NIT_Z1_QC}
		\osc{Z}_k^{(1)}  = & \sum_{\kappa \neq 0} \frac{ i e^{ i \kappa \chi}}{\kappa}  \Biggl[ -  \PD{\nit{s}_k^{(0)}}{\nit{P}_j} \osc{Y}_{j, \kappa}^{(1)} +  \sum_{\kappa' \neq 0}  \PD{\osc{Z}_{k, \kappa'}^{(0)}}{\nit{P}_j} \osc{F}_{j,\kappa - \kappa'}   \Biggr].
	\end{split}
\end{align}
\end{subequations}
Compared with the 1PA terms of \PaperI, all contributions proportional to $\osc{f}_w$ and $\PD{}{\nit{\zeta}}$ are absent, since the forcing functions no longer depend on an evolving phase. As in \PaperI, we include the sub-leading transformations $\osc{Y}_j^{(2)}$ and $\osc{Z}_k^{(1)}$ in the ``1PA model.''

\subsection{Second postadiabatic terms}
To reach acceptable accuracy for comparable-mass systems, we again extend the calculation to second postadiabatic order (2PA). The resulting averaged terms are
\begin{subequations} \label{eq:2PA_Averaged_Terms_QC}
\begin{align}
	\begin{split}
		\nit{F}_j^{(2)} = & \avg{\PD{\osc{Y}_j^{(2)}}{\nit{P}_k} \osc{F}_{k} }  - \frac{1}{2} \frac{\partial^2 \nit{F}_j^{(0)}}{\partial \nit{P}_k \partial \nit{P}_i}  \avg{ \osc{Y}_k^{(1)} \osc{ Y}_i^{(1)} } ,
	\end{split}\\
	\begin{split}
		\nit{s}_k^{(2)} = & \avg{\PD{\osc{Z}_k^{(1)}}{\nit{P}_j} \osc{F}_{j} } - \frac{1}{2} \frac{\partial^2 \nit{s}_k^{(0)}}{\partial \nit{P}_i \partial \nit{P}_j}  \avg{ \osc{Y}_i^{(1)} \osc{ Y}_j^{(1)} } .
	\end{split}
\end{align}
\end{subequations}
We refer to the model that adds these 2PA contributions to the 1PA model as the ``2PA model.''

Unlike the averaged equations of \PaperI, which become ill-behaved as $e \rightarrow 0$, the above averaged equations are valid for any value of the eccentricity, including the quasicircular limit. The price is the enlarged intrinsic parameter space $(p,\alpha,\beta)$ discussed in Sec.~\ref{section:OOE_QC}.

\subsection{Summary of results} \label{section:NIT_Summary_QC}
Combining the above results, we obtain the averaged equations of motion for the NIT variables in the $(\alpha,\beta)$ parametrization through second postadiabatic order:
\begin{subequations} \label{eq:NIT_EoM_QC_Explicit}
\begin{align}	
	\begin{split}
		\frac{d \nit{p}}{d \chi} &=  \sum_{n =0}^{2}\epsilon^{n+1} \nit{F}_p^{(n)}(\nit{p},\nit{\alpha},\nit{\beta}) + \HOT{4} , 
	\end{split} \\
	\begin{split}
		\frac{d \nit{\alpha}}{d \chi} &=  \sum_{n =0}^{2}\epsilon^{n+1} \nit{F}_\alpha^{(n)}(\nit{p},\nit{\alpha},\nit{\beta}) + \HOT{4}, 
	\end{split}\\
	\begin{split}
		\frac{d \nit{\beta}}{d \chi} &= \sum_{n =0}^{2}\epsilon^{n+1} \nit{F}_\beta^{(n)}(\nit{p},\nit{\alpha},\nit{\beta}) + \HOT{4}, 
	\end{split} \\
	\begin{split}
		\frac{d \nit{t}}{d \chi} & =  \sum_{n =0}^{2}\epsilon^n \nit{f}_t^{(n)}(\nit{p},\nit{\alpha},\nit{\beta}) + \HOT{3} ,
	\end{split}\\
	\begin{split}
		\frac{d \nit{\phi}}{d \chi} & = \sum_{n =0}^{2}\epsilon^n \nit{f}_\phi^{(n)}(\nit{p},\nit{\alpha},\nit{\beta}) + \HOT{3} 
	\end{split}
\end{align}
\end{subequations}
The NIT variables are related to the OOE variables, which are needed to set consistent initial conditions, via
\begin{subequations} \label{eq:OOE_NIT_Initial_Conditions_QC}
\begin{align}
	\begin{split}
		\nit{p} &= p + \sum_{n =1}^{2}\epsilon^n Y_p^{(n)}(p,\alpha,\beta,\chi)  + \HOT{3} , 
	\end{split} \\
	\begin{split}
		\nit{\alpha} &=  \alpha + \sum_{n =1}^{2}\epsilon^n Y_\alpha^{(n)}(p,\alpha,\beta,\chi) + \HOT{3},
	\end{split}\\
	\begin{split}
		\nit{\beta} &=  \beta + \sum_{n =1}^{2}\epsilon^n Y_\beta^{(n)}(p,\alpha,\beta,\chi) + \HOT{3},
	\end{split}\\
	\begin{split}
		\nit{t} & = t +  \sum_{n =0}^{1}\epsilon^n Z_t^{(n)}(p,\alpha,\beta,\chi)  +  \HOT{2},
	\end{split}\\
	\begin{split}
		\nit{\phi} & = \phi + \sum_{n =0}^{1}\epsilon^n Z_\phi^{(n)}(p,\alpha,\beta,\chi) +  \HOT{2}.
	\end{split}
\end{align}
\end{subequations}
The OOE variables can be recovered from the solutions to the NIT equations of motion using the inverse transformations of Eqs.~\eqref{eq:inverse_transformation_QC}. In both parametrizations, the solutions are functions of $\chi$, so one must interpolate $\nit{t}(\chi)$ in order to express them as functions of time.

\subsection{Reduction of the parameter space} \label{section:ParameterSpaceReduction}
At first sight, the $(\alpha,\beta)$ parametrization requires every forcing function, transformation function and averaged term to be tabulated over the four-dimensional space $(\nu,p,\alpha,\beta)$. The extra dimension, however, is a rotation, and its effect can be worked out analytically. To see this, we combine the two components of the eccentricity vector into the complex variable
\begin{equation} \label{eq:EccVector}
	\En \equiv \beta + i \alpha = e\, e^{i w},
\end{equation}
in terms of which Eq.~\eqref{eq:r_of_p_alpha_beta} reads $r = pM/[1 + \mathrm{Re}(\En\, e^{-i\chi})]$. Rotating the eccentricity vector by an angle $\theta$ while shifting the radial phase by the same angle, $\En \rightarrow e^{i\theta} \En$ and $\chi \rightarrow \chi + \theta$, describes the same orbit. The EOB variables $(r,p_{r_*},p_\phi)$, and hence every function built from them, are therefore covariant under this joint rotation. Scalar quantities such as $F_p$, $f_t$, $f_\phi$ and the waveform mode amplitudes are invariant, while the forcing of the eccentricity vector,
\begin{equation} \label{eq:F_EccVector}
	F_{\En} \equiv F_\beta + i F_\alpha = e^{i w} \left( F_e - i\, e f_w \right),
\end{equation}
rotates with it. The same holds for the corresponding transformation functions $Y_\En^{(n)} = Y_\beta^{(n)} + i Y_\alpha^{(n)}$ and averaged terms $\nit{F}_\En^{(n)}$. Assigning a spin weight $s$ to each function, with $s = 0$ for scalars and $s = 1$ for the vector quantities, the Fourier coefficients of Eq.~\eqref{eq:Fourier_QC} obey
\begin{equation} \label{eq:FourierCovariance}
	A_\kappa(\nu,p,\alpha,\beta) = e^{i (s - \kappa) w} A_\kappa(\nu,p,e,0) .
\end{equation}
The entire dependence on $w$ is thus a phase, and every function is determined by its values on the slice $w = 0$, which is precisely the $(\nu,p,e)$ parameter space of \PaperI. On this slice $\zeta = \chi$, so Eq.~\eqref{eq:F_EccVector} relates the new coefficients directly to those of \PaperI, and the Fourier data computed there can be reused. The averaged equation of motion for the eccentricity vector takes the compact form
\begin{equation} \label{eq:NIT_EoM_EccVector}
	\frac{d \nit{\En}}{d\chi} = \left[ \sp\, g^{(0)}(\nu,\nit{p},\nit{e}^2) + \sp^2 g^{(1)}(\nu,\nit{p},\nit{e}^2) + \sp^3 g^{(2)}(\nu,\nit{p},\nit{e}^2) \right] \nit{\En},
\end{equation}
with complex scalar coefficients $g^{(n)}$ that remain finite as $\nit{e} \rightarrow 0$.

The covariance also tells us how each coefficient behaves in the circular-orbit limit. Writing $m = s - \kappa$, smoothness in $(\alpha,\beta)$ requires
\begin{align} \label{eq:ReducedCoefficients}
	\begin{split}
	A_\kappa(\nu,p,\alpha,\beta) & = \En^{m}\, a_\kappa(\nu,p,e^2)  \quad (m \geq 0), \\
	A_\kappa(\nu,p,\alpha,\beta) & = \bar{\En}^{|m|} a_\kappa(\nu,p,e^2) \quad (m < 0),
	\end{split}
\end{align}
where the $a_\kappa$ are smooth functions of $e^2$. Each coefficient therefore vanishes as $e^{|m|}$, and we interpolate the reduced coefficients $a_\kappa$ as functions of $(\nu,p,e^2)$ rather than the $A_\kappa$ themselves. The derivatives with respect to $\alpha$ and $\beta$ that enter the second-order terms then follow from the analytic prefactors and the $e^2$ derivatives of the $a_\kappa$. Details of the grid, of the reduction at high harmonics and of the small-eccentricity data are given in Sec.~\ref{section:InterpolationGrid}.

In summary, the $(\alpha,\beta)$ formalism is tabulated on the same three-dimensional $(\nu,p,e)$ grid as \PaperI, extended by nodes below $e = 0.01$ and by the analytic circular-orbit limit at $e = 0$. It is valid for all $0 \leq e < 1$, and we use it for the entire inspiral without any transition between parametrizations.

\section{Waveform Generation} \label{section:Waveforms}
We now discuss waveform generation from the averaged dynamics. We first recall the waveform-mode expressions and the sampling method of \PaperI, and then introduce a Fourier decomposition of the waveform modes with respect to the radial phase that removes the need to evaluate the full mode expressions at every time step.

\subsection{Waveform modes} \label{section:WaveformModes}
The waveform strain is decomposed into $-2$ spin-weighted spherical-harmonic modes as
\begin{equation} \label{eq:StrainDecomposition}
	h = h_+ - i h_\times = \sum_{\ell \geq 2} \sum_{m=-\ell}^{\ell} h_{\ell m} \,_{-2}Y_{\ell m}(\theta, \phi).
\end{equation}
As in \PaperI, we employ the 3PN instantaneous contributions to the modes from Ref.~\cite{Gamboa:2024imd}, which take the general form
\begin{equation} \label{eq:WaveformMode}
	h_{\ell m} =  \mathcal{A}_{\ell m} \left(r,p_r,p_\phi \right) \exp \left( - i m \phi \right),
\end{equation}
and we model only the dominant $(\ell,m) = (2,2)$ mode. Explicit expressions for $\mathcal{A}_{22}$ up to 3PN order for nonspinning systems are given in Appendix~D of \PaperI.

\subsection{Waveform generation in \PaperI} \label{section:WaveformsNIT}
Solving the NIT equations of motion yields the NIT variables on sparse, non-uniform steps in $\chi$, which are too sparse to resolve the orbital-timescale oscillations of the waveform. In \PaperI, we therefore sampled the NIT solutions onto a uniform grid in $\chi$ with spacing $\Delta \chi$ set by the Nyquist criterion for the highest Fourier mode present in the leading-order transformation $Z_t^{(0)}$, reconstructed the EOB variables $(r,p_{r_*},\phi,p_\phi)$ at each sample point from the relations of Sec.~\ref{section:OOE_Summary}, and evaluated the mode expression \eqref{eq:WaveformMode}. The modes were then interpolated and upsampled onto a dense uniform time grid that can be Fourier transformed for mismatch and likelihood calculations. In that reconstruction the intrinsic NIT variables $(\nit{p},\nit{e},\nit{\zeta})$ were used in place of the osculating elements $(p,e,\zeta)$, which spares the inverse transformation at every sample point but, as noted in \PaperI, leaves a mismatch floor of $\sim 10^{-6}$. In Sec.~\ref{section:WaveformAssembly} we remove this floor by interpolating the modes directly as functions of the NIT variables.

Reconstructing the EOB variables and evaluating the 3PN mode amplitudes at each of the many sample points nevertheless dominated the total waveform-generation time in \PaperI, limiting the overall speed-up to $1.5$--$8\times$ despite a speed-up of the dynamics of up to two orders of magnitude.

\subsection{Fourier decomposition of the waveform modes} \label{section:WaveformFourierDecomposition}
To remove this bottleneck, we note that the mode amplitudes $\mathcal{A}_{\ell m}$, when expressed in terms of the osculating orbital elements, are periodic functions of the radial phase. In the parametrization of \PaperI, we can therefore write
\begin{equation} \label{eq:WaveformFourier_Ecc}
	h_{\ell m} =\sum_{n = -n_\text{max}}^{n_\text{max}} \mathcal{A}_{\ell m n}(\nu, p,e) \exp \left[ i \left( n \zeta  -  m \phi \right) \right],
\end{equation}
where the Fourier coefficients $\mathcal{A}_{\ell m n}$ are smooth functions across the $(\nu,p,e)$ parameter space, the sum is truncated at $|n| \leq n_\text{max}$, and the sign of the exponent follows the convention of Eq.~\eqref{eq:Fourier_QC}. In the $(\alpha,\beta)$ parametrization, the corresponding decomposition is with respect to $\chi$:
\begin{equation} \label{eq:WaveformFourier_QC}
	h_{\ell m} =\sum_{n = -n_\text{max}}^{n_\text{max}} \mathcal{A}_{\ell m n}(\nu, p,\alpha,\beta) \exp \left[ i \left( n \chi  -  m \phi \right) \right].
\end{equation}
The mode amplitudes are scalars under the joint rotation of Sec.~\ref{section:ParameterSpaceReduction}, so their coefficients obey the covariance \eqref{eq:FourierCovariance} with spin weight $s = 0$, that is $\mathcal{A}_{\ell m n}(\nu,p,\alpha,\beta) = e^{-i n w} \mathcal{A}_{\ell m n}(\nu,p,e)$, and they are tabulated on the same three-dimensional grid and with the same reduction near the circular-orbit limit as the forcing functions. They are computed once by sampling the mode amplitude over a radial period and applying an FFT, in the same manner as the forcing functions in Sec.~\ref{section:FourierDecomposition_QC}. At each sample point along the inspiral, the waveform mode is then obtained from a short sum of interpolated coefficients multiplied by phase factors, rather than by reconstructing the EOB variables and evaluating the full 3PN mode expression. The number of harmonics needed grows with the eccentricity, and we choose $n_\text{max}$ adaptively for each inspiral as described in Sec.~\ref{section:WaveformSettings}. We examine the convergence with $n_\text{max}$ in Sec.~\ref{section:2PAOrders}.

We also find that the accuracy of the mode amplitudes has a much weaker effect on the mismatch than the accuracy of the phasing, which is already accurately determined by the 2PA NIT dynamics. Modest errors in the interpolated coefficients are therefore tolerable, and we quantify the effect of the truncation settings in Sec.~\ref{section:WaveformSettings}. Finally, Eqs.~\eqref{eq:WaveformFourier_Ecc} and \eqref{eq:WaveformFourier_QC} have the same structure as the multi-voice decomposition of the waveform used in the \texttt{FastEMRIWaveforms} package \cite{Chua:2020stf,Katz:2021yft,Speri:2023jte,Chapman-Bird:2025xtd,Hughes:2021exa}. 

However, our approach differs in three important ways. First, the multi-voice decomposition represents the waveform in terms of Boyer-Lindquist time action angles which advance uniformly in coordinate time \cite{Hughes:2021exa,Lynch:2024hco}. Sampling in the radial action angle, also known as the mean anomaly, the number of radial harmonics required grows steeply with eccentricity, reaching $\sim 50 $ by $e \simeq 0.7$ \cite{Chapman-Bird:2025xtd}. Our expansion is in the relativistic anomaly, which absorbs the nonuniform sweep of the orbit through periastron. As a result, its Fourier series converges much faster. We find that only five harmonics suffice at $e_0 = 0.7$ (Sec.~\ref{section:2PAOrders}).

 Second, the EMRI waveform amplitudes of the multi-voice are typically sampled as functions of the adiabatic variables, whereas ours are tabulated at the NIT variables, which include the oscillatory part of the transformation (Sec.~\ref{section:WaveformAssembly}).

 Finally, our dynamics are computed as a function of the relativistic anomaly instead of time. This means we require an extra step to convert the waveform to a uniform time grid, which is done by inverting the relation $t(\chi)$ (Sec.~\ref{section:WaveformAssembly}). This is not needed in the standard multi-voice decomposition, where the waveform is assembled from dynamics and amplitudes that vary slowly with Boyer-Lindquist time. 

One similarity these two approaches share is that the sum over $n$ at each time step is embarrassingly parallel and can be evaluated efficiently on both CPUs and GPUs. The speed-up obtained by modest parallelization over 8 CPU cores is examined in Sec.~\ref{section:WaveformSpeedup}.

\subsection{Assembly of the waveform on the time grid} \label{section:WaveformAssembly}
Two further ingredients are needed to turn Eq.~\eqref{eq:WaveformFourier_QC} into an efficient generator.

The first is the choice of variables at which the mode coefficients are tabulated. The averaged dynamics provide the NIT variables $\nit{P}_j(\chi)$, whereas the coefficients of Eq.~\eqref{eq:WaveformFourier_QC} are functions of the osculating elements, which differ from the NIT variables by the oscillatory transformation \eqref{eq:transformation_QC}. If one simply evaluates $\mathcal{A}_{\ell m n}$ at the NIT variables, the resulting $\mathcal{O}(\sp\, Y^{(1)})$ error in the amplitude caps the mismatch at $\sim 10^{-6}$--$10^{-7}$ regardless of the accuracy of the phasing. This floor is not intrinsic to the tabulated data. The OOE model of Sec.~\ref{section:OOEModel}, which integrates the un-averaged equations with the same interpolated coefficients, reaches mismatches about two orders of magnitude lower. We therefore interpolate the Fourier coefficients of the mode amplitude as functions of the NIT variables, $\nit{\mathcal{A}}_{\ell m n}(\nu,\nit{p},\nit{e})$. At each node of the grid, we map the NIT variables to the osculating elements with the inverse transformation \eqref{eq:inverse_transformation_QC} along one radial period, evaluate the EOB variables and the mode amplitude there, and take the FFT. Since $\epsilon$ is a bookkeeping parameter that is set to one, the inverse transformation to next-to-leading order can be substituted directly. Note that we use this construction for all NIT models when we should only use the leading order inverse transformation for the 0PA model. However, this minor discrepancy in the amplitudes is subdominant to the error in the phasing caused by missing PA orders in the dynamics. The waveform then follows from the NIT solution with no inverse transformation required. These coefficients obey the same rotational covariance as before, so they are stored on the same three-dimensional grid.

The second ingredient is the change of independent variable from $\chi$ to time. Along the inspiral, the physical time and azimuth follow from the inverse transformation \eqref{eq:inverse_transformation_QC_S},
\begin{subequations} \label{eq:tphi_of_chi}
\begin{align}
	t(\chi) &= \nit{t}(\chi) - \sum_{\kappa} C^t_\kappa(\chi)\, e^{i \kappa \chi}, \\
	\phi(\chi) &= \nit{\phi}(\chi) - \sum_{\kappa} C^\phi_\kappa(\chi)\, e^{i \kappa \chi},
\end{align}
\end{subequations}
where $C^t_\kappa = Z^{(0)}_{t,\kappa} + \sp \nit{Z}^{(1)}_{t,\kappa}$, and likewise for $\phi$, are evaluated at the NIT variables and therefore vary only on the radiation-reaction timescale. The waveform is required on a uniform time grid $t_j = t_0 + j \Delta t$. Note that the grid does not have to be uniform as each sample is inverted independently, so the waveform can be generated on any set of times specified by the user. We use a uniform grid only because our mismatch calculation requires one.

For each sample we solve $t(\chi_j) = t_j$ for $\chi_j$, and then evaluate $\phi(\chi_j)$ and the mode sum \eqref{eq:WaveformFourier_QC}. We do this in three steps that mirror the three timescales of the problem. First, the coefficients $C^t_\kappa$, $C^\phi_\kappa$ and $\nit{\mathcal{A}}_{\ell m n}$ are evaluated only at the nodes of the numerical solution of the averaged equations, which number a few hundred per inspiral. Second, they are interpolated onto a coarse guess grid in $\chi$ with $24$ points per radial period, on which $t(\chi)$ is evaluated from Eq.~\eqref{eq:tphi_of_chi}. Since $t(\chi)$ is monotonic, this tabulates the inverse function $\chi(t)$ on the same grid, and interpolating it, by cubic Hermite interpolation with the exact slopes $d\chi/dt$, at every sample time $t_j$ provides the initial guess $\chi_j^{(0)}$. Third, each guess is refined by Newton--Raphson iteration on the root of $t(\chi) - t_j$, i.e.,
\begin{equation} \label{eq:NewtonStep}
	\chi_j^{(k+1)} = \chi_j^{(k)} - \frac{t(\chi_j^{(k)}) - t_j}{dt/d\chi\,(\chi_j^{(k)})},
\end{equation}
where
\begin{equation}
	\frac{dt}{d\chi} \simeq \frac{d\nit{t}}{d\chi} - \sum_{\kappa} i \kappa\, C^t_\kappa\, e^{i \kappa \chi},
\end{equation}
$d\nit{t}/d\chi$ is the averaged right-hand side of Eq.~\eqref{eq:transformed_EoM_QC} and the $\chi$ derivative of the slowly varying coefficients $C^t_\kappa$ is neglected. Two iterations suffice to reduce the residual $|t(\chi_j) - t_j|$ below $10^{-7} M$ even at $e_0 = 0.85$. After these two iterations, $\phi(\chi_j)$ and the mode sum are evaluated at the converged $\chi_j$. Since $t(\chi)$ is monotonic and the guess grid resolves its periodic modulation, the iteration converges in all but a handful of samples per waveform, which are identified by their residual and bisected instead.

The work per output sample is thus a small, fixed number of arithmetic operations, with no interpolant calls and no evaluation of the 3PN mode expression, and the samples are mutually independent. The assembly can therefore be distributed over blocks of the time grid, with only the guess-grid data shared between them, allowing for efficient parallelization.

The late inspiral after the handover is generated as in \PaperI, by integrating the EOB equations from the inverse-transformed variables at the handover point and evaluating the mode expression on the remaining time samples. The numerical settings of the assembly are given in Sec.~\ref{section:WaveformSettings}.


\section{Implementation}\label{section:Implementation}
We now outline the implementation of our Mathematica code for solving the dynamics and generating waveforms. We describe the interpolation grid and the Fourier data tabulated on it, the initial conditions, the numerical integration and the handover to the EOB dynamics, the settings of the waveform generator, the OOE model used as a numerical reference, and the mismatch calculation.

\subsection{Interpolation grid} \label{section:InterpolationGrid}
As discussed in Sec.~\ref{section:ParameterSpaceReduction}, all forcing functions, transformation functions, averaged terms and waveform modes are tabulated on a three-dimensional grid in $(\nu, p, e)$ and evaluated by interpolation. We use the grid of \PaperI, extended towards circular orbits. It is uniformly spaced in $\nu$, in $u = \log(p^{-3})$ and in $e$, given by
\begin{itemize}
	\item $\nu \in [0.01, 0.25]$ with $\Delta \nu = 0.01$,
	\item $u \in [-17, \log(7.9^{-3})]$ (which corresponds to $p \in [7.9, 289.069]$) with $\Delta u = 0.05$,
	\item $e \in [0.01, 0.9]$ with $\Delta e = 0.01$, supplemented by the nodes $e = 0$, $0.0025$, $0.005$ and $0.0075$.
\end{itemize}
This results in a grid of $25 \times 217 \times 94 = 509{,}950$ points, on which we store the Fourier coefficients of the forcing functions $F_p$, $F_\En$, $f_t$ and $f_\phi$, of the second-order transformations $Y^{(2)}_j$ and $Z^{(1)}_k$ and their inverses, and of the waveform modes, with up to $\pm 30$ radial harmonics each.

For $e \geq 0.06$, the Fourier coefficients are obtained from those of \PaperI\ through Eq.~\eqref{eq:F_EccVector}. For smaller eccentricities, the double-precision FFT of \PaperI\ is not accurate enough, since its absolute error floor of order $10^{-10}$ becomes a few percent of the small harmonics at $e = 0.02$. All nodes with $e \leq 0.05$ were therefore recomputed in $32$-digit arithmetic, and the $e = 0$ row is given by the analytic circular-orbit limit of Appendix~\ref{appendix:OOE_EoM_QC}, except for the reduced coefficients with $|m| \geq 1$, whose $e = 0$ row is obtained by quadratic extrapolation in $e^2$ from the three smallest nodes.

The reduced coefficients of Eq.~\eqref{eq:ReducedCoefficients} are interpolated by cubic Hermite interpolation in $(\nu, p, e^2)$, and the real averaged terms by cubic splines. For high harmonics, dividing by $e^{|m|}$ amplifies the numerical noise in the tabulated coefficients, so we cap the power that is factored out at $|m| \leq 5$ for the forcing and transformation functions and at $|m| \leq 3$ for the waveform modes. The remaining factor of $e^{|m| - 5}$ (or $e^{|m| - 3}$) is interpolated along with the coefficient, which is harmless since these harmonics are negligible at the small eccentricities where the reduction matters. We keep harmonics up to $|\kappa| \leq 30$ for the forcing and leading-order transformation functions and $|\kappa| \leq 15$ for the second-order and inverse terms. The waveform-mode coefficients at the NIT variables (Sec.~\ref{section:WaveformAssembly}) are computed from $128$ samples per orbit using the 2PA inverse transformation. The same tables are used for the waveforms of all three PA models, so that the comparison between them isolates the accuracy of the dynamics.

Building these tables from the Fourier data of \PaperI\ took about eight hours on six to eight kernels of an Apple M1 Max, of which about three hours were spent on the $32$-digit recomputation. The tables occupy $19$\,GB on disk, including the checkpoints of the build, and $4.6$\,GB in memory. This is a one-off cost per model, and the grid is deliberately dense, so that interpolation is not the leading source of error, rather than optimized for size.

\subsection{Initial conditions} \label{section:InitialConditions}
As in \PaperI, we specify the initial conditions of the OOE system and derive those of the EOB and NIT systems from them, so that all three describe the same physical inspiral. The OOE initial conditions are now $(p_0, \alpha_0, \beta_0, \phi_0)$ at $t_0 = 0$.
 For a given initial eccentricity $e_0$ and initial relativistic anomaly $\zeta_0 = \chi_0 - w_0$, we set $\chi_0 = 0$ and obtain $(\alpha_0, \beta_0) = e_0 (\sin w_0, \cos w_0)$ with $w_0 = -\zeta_0$ from Eq.~\eqref{eq:alpha_beta_def}. By the rotational covariance of Sec.~\ref{section:ParameterSpaceReduction}, this is equivalent to any other choice of $\chi_0$ with the same $\zeta_0$. For continuity with \PaperI, all runs in this work start at periastron, $\zeta_0 = 0$, i.e. $\alpha_0 = 0$ and $\beta_0 = e_0$, although the method applies unchanged to any choice of $\zeta_0$. Rather than $p_0$, we specify the initial orbit-averaged gravitational-wave frequency $f_0$ (i.e., twice the orbit-averaged orbital frequency) and root-find numerically for $p_0$ using
\begin{equation}
	f_0 = \frac{1}{\pi} \avg{\Omega_\phi} = \frac{1}{\pi} \left( \frac{\nit{f}_\phi^{(0)}(\nu, p_0, e_0)}{\nit{f}_t^{(0)}(\nu, p_0, e_0)} \right),
\end{equation}
where $\nit{f}_t^{(0)}$ and $\nit{f}_\phi^{(0)}$ are the leading-order terms of Eq.~\eqref{eq:NIT_EoM_QC_Explicit}, which depend on $\alpha$ and $\beta$ only through $e$. The initial conditions of the EOB system $(r_0, p_{r_* 0}, \phi_0, p_{\phi 0})$ then follow from Eq.~\eqref{eq:r_of_p_alpha_beta} and the relations of \PaperI\ for $p_\phi$ and $p_{r_*}$, and those of the NIT system from the transformation \eqref{eq:OOE_NIT_Initial_Conditions_QC} evaluated at $\chi_0$, truncated at the postadiabatic order of the model.

\subsection{Numerical integration and handover} \label{section:Handover}
The NIT equations of motion \eqref{eq:NIT_EoM_QC_Explicit} are solved with the \texttt{NDSolve} function in Mathematica with the method set to \texttt{ExplicitRungeKutta} with \texttt{DifferenceOrder} $5$ and \texttt{AccuracyGoal} and \texttt{PrecisionGoal} set to $10$. The averaged dynamics take tens of milliseconds with any method. The choice matters because the integration steps form the sparse $\chi$ grid on which the waveform coefficients of Sec.~\ref{section:WaveformAssembly} are evaluated, and this method resolves them with roughly a third of the steps of the default method at equal accuracy. The EOB dynamics, both for the reference waveforms and for the late inspiral, are integrated as in \PaperI\ using RK4 with \texttt{AccuracyGoal} $9$ and \texttt{PrecisionGoal} $\infty$.

The handover from the NIT to the EOB dynamics follows \PaperI. At each step we check whether
\begin{itemize}
	\item $\nit{p} < 8.5$,
	\item $\Q < 100$,
\end{itemize}
where the adiabaticity parameter $\Q = \avg{\Omega_\phi}^2 / \avg{\dot{\Omega}_\phi}$ is estimated from the adiabatic terms as in \PaperI. If either condition is violated, we transform back to the OOE variables using Eqs.~\eqref{eq:inverse_transformation_QC}, construct the EOB variables and integrate the EOB equations of motion to the end of the evolution at $r = 4M$. Throughout this paper, we refer to this final EOB segment, from the handover to the end of the evolution, as the late inspiral. The third condition of \PaperI, which handed over to the EOB dynamics once $\nit{e} < 0.035$, is no longer needed since the $(\alpha,\beta)$ equations are regular at $e = 0$. Systems whose initial adiabaticity parameter satisfies $\Q < 105$ would start inside or at the edge of the handover region, and are excluded from the comparisons of Sec.~\ref{section:Results}, as in \PaperI.

\subsection{Waveform generation} \label{section:WaveformSettings}
The waveform is assembled as described in Sec.~\ref{section:WaveformAssembly}. We keep the harmonics of the transformation functions whose coefficients exceed $10^{-6}$ of the leading one and the modes whose amplitude exceeds $10^{-6}$ of the circular $n = 0$ mode. The inversion of $t(\chi)$ uses $24$ guess points per orbit and two Newton iterations with a bisection safeguard. The samples are assembled in blocks of $2 \times 10^5$ using vectorized array operations, or with an equivalent compiled kernel when the blocks are distributed over several processor cores. Relative to the most conservative settings, these choices save about a third of the walltime and change the mismatches by at most a factor of two at the $10^{-8}$ level. The cost of a waveform is then a fixed $0.2$--$0.25$\,s for the averaged dynamics, the evaluation of the coefficients at its nodes and the late inspiral, plus about $1.6$\,$\mu$s per sample.

\subsection{OOE model} \label{section:OOEModel}
To separate the error of the tabulated data from that of the postadiabatic truncation, we also integrate the un-averaged osculating-element equations \eqref{eq:OOE_EoM_QC} with forcing functions reconstructed from interpolated Fourier coefficients, i.e., the OOE model of \PaperI. Its mismatch against the EOB waveform measures the error of the Fourier truncation and interpolation alone and so sets a numerical floor for the NIT waveforms. For this model, the forcing functions are represented as real cosine and sine series in $\chi - w$ whose reduced coefficients are interpolated by cubic splines in $(\nu, p, e)$, as in \PaperI. Between the grid nodes this representation is about ten times more accurate than the Hermite tables used for the averaged dynamics, and so provides a slightly lower floor. The waveform is generated as in \PaperI\ from the EOB variables sampled at $256$ points per orbit in $\chi$.

\subsection{Mismatch calculation} \label{section:MismatchCalculation}
Mismatches are computed as in \PaperI\ using the \texttt{GWMMatch} package \cite{lynch_2026_20426391}. We take the $(2,2)$ mode of the waveform, sampled at $4096$\,Hz from an initial orbit-averaged gravitational-wave frequency of $10$\,Hz, apply a Planck taper to both ends \cite{McKechan:2010kp}, zero-pad to a common length and compute the noise-weighted inner product with the \texttt{A+} LIGO design sensitivity for O5 \cite{LIGO-T0900288-v3} with a low-frequency cutoff of $10$\,Hz. The mismatch $\mathcal{M}_{22}$ is then one minus the faithfulness maximized over time and phase shifts. Since this paper is concerned with the accuracy of the NIT dynamics rather than with the physical accuracy of the waveform model, we do not consider other detectors or higher modes here.

\section{Results} \label{section:Results}
We now assess the accuracy and speed of the NIT dynamics and the fast waveform generation against the full EOB evolution. Throughout, the reference waveform is generated with the EOB dynamics and the recipe of \PaperI\ (Sec.~\ref{section:Implementation}), the NIT waveforms with the default settings of Sec.~\ref{section:WaveformSettings}, and mismatches are computed as in \PaperI\ with the O5 design sensitivity curve and a low-frequency cutoff of 10\,Hz. Unless stated otherwise, all systems start at periastron with an orbit-averaged gravitational-wave frequency of $f_0 = 10$\,Hz.

\subsection{Case Study} \label{section:CaseStudy}
\begin{figure*}[tbp]
\includegraphics[width=\textwidth]{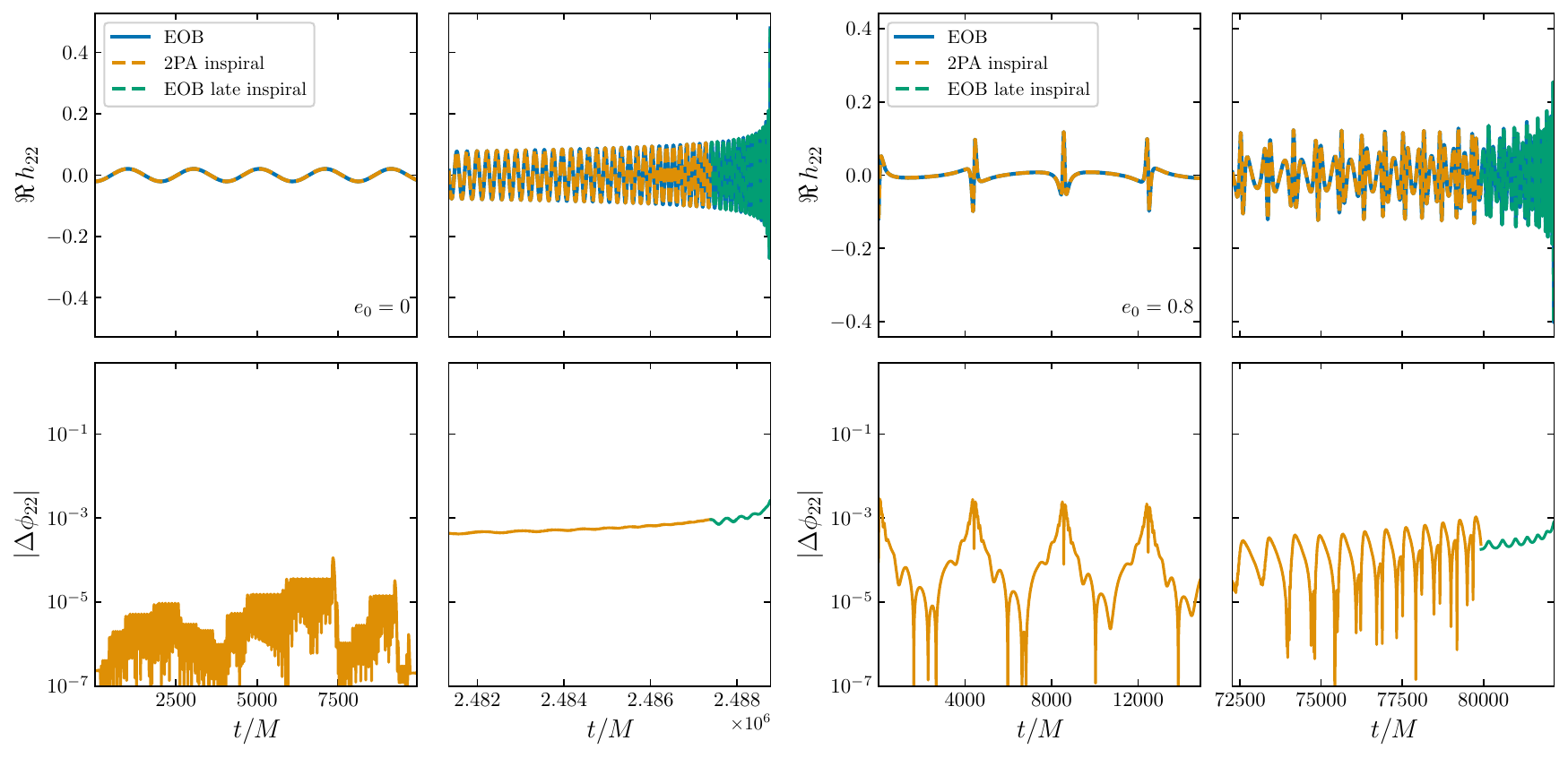}
\caption{
The $(\ell,m) = (2,2)$ mode of the EOB and 2PA NIT waveforms for two binaries with $M = 10 M_\odot$ and $q = 1$ starting at an orbit-averaged frequency of $10$\,Hz: a quasicircular inspiral with $e_{10 \rm Hz} = 0$ (left block, $p_0 = 74.8$, $123$\,s of signal) and a highly eccentric one with $e_{10 \rm Hz} = 0.8$ (right block, $p_0 = 28.3$, $4$\,s). In each block, the top panels show the real part of the mode and the bottom panels the phase difference between the EOB and NIT waveforms, for the early inspiral (left) and the late inspiral (right), with a gap in time between the two. The late-inspiral panels highlight where the NIT dynamics (dashed orange) hand over to the EOB dynamics (dashed green). The phase difference is accumulated throughout the full inspiral.
\label{fig:CaseStudyWaveforms}
}
\end{figure*}
We first illustrate the method on two binaries with $M = 10 M_\odot$ and $q = 1$ that bracket the eccentricities considered in this work (Fig.~\ref{fig:CaseStudyWaveforms}). The first is a quasicircular inspiral, $e_0 = 0$, which is precisely the regime that the parametrization of \PaperI\ could not treat, and which at this low total mass produces $123$\,s of signal in band and $\sim 10^3$ orbits. The second has $e_0 = 0.8$, the largest eccentricity for which this system starts outside the handover margin $\mathcal{Q} \geq 105$ of Sec.~\ref{section:Handover}, and spends only $4$\,s in band. In both cases the 2PA NIT waveform is visually indistinguishable from the EOB one throughout the inspiral. The phase difference grows secularly from $10^{-6}$ to $9 \times 10^{-4}$\,rad over the quasicircular inspiral and reaches $2.6 \times 10^{-3}$\,rad at the end of the late inspiral. For the eccentric binary it oscillates with the periastron passages at the $10^{-4}$--$3 \times 10^{-3}$\,rad level. The corresponding mismatches are $3.6 \times 10^{-9}$ for the quasicircular binary and $9.4 \times 10^{-8}$ for the eccentric one. The quasicircular waveform takes $13$\,s to generate with the EOB dynamics and $1.1$\,s with the 2PA NIT, a speed-up of $12\times$, while the eccentric waveform takes $8.4$\,s and $0.47$\,s respectively, a speed-up of $18\times$ (see Sec.~\ref{section:WaveformSpeedup} for the timing setup).

\subsection{Comparison of postadiabatic orders} \label{section:PAOrders}

\begin{figure*}[tbp]
	\includegraphics[width=.9\textwidth]{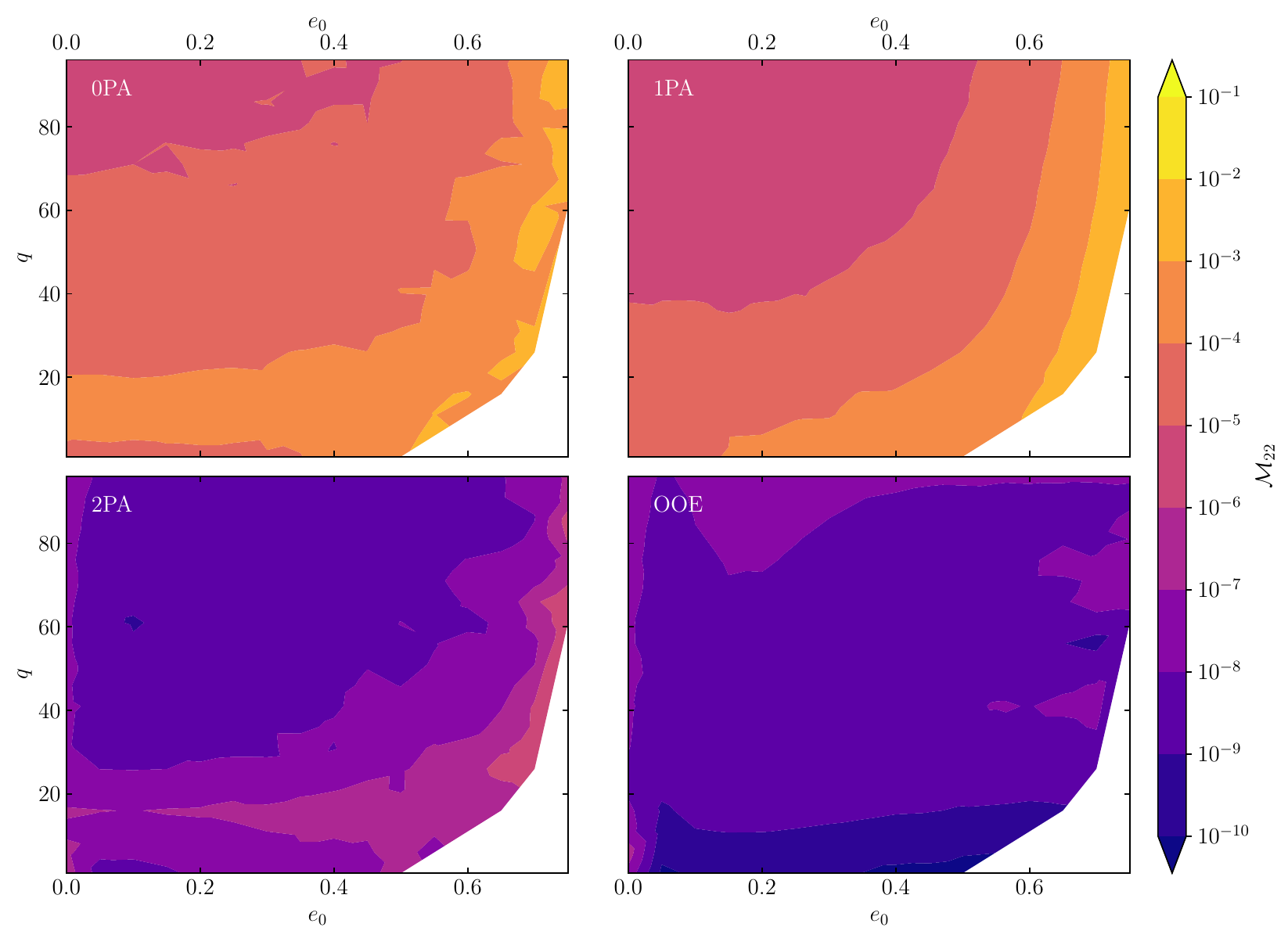}
	\caption{Contour plot of the mismatch between the EOB waveform and the OOE and NIT waveforms at different postadiabatic orders for $M = 60 M_\odot$ over $q \in [1,96]$ and $e_0 \in [0, 0.85]$, starting with an orbit-averaged frequency $f_0 = 10$\,Hz. Mismatches use the O5 design sensitivity curve and a low-frequency cutoff of 10\,Hz. White regions are systems that begin after the NIT handover conditions are already satisfied.
		\label{fig:Model_Mismatch_Contours}
	}
\end{figure*}

\begin{figure}[tbp]
	\includegraphics[width=1.0\columnwidth]{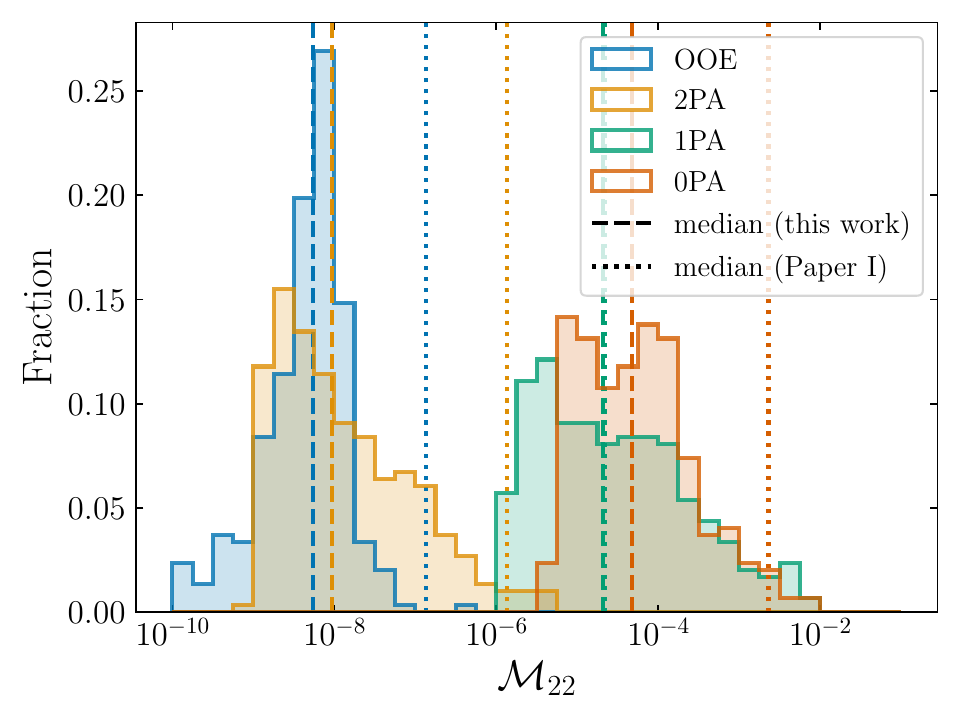}
	\caption{Histogram of the mismatch between the EOB waveform and the OOE and NIT waveforms at different postadiabatic orders for $M = 60 M_\odot$ over $q \in [1,96]$ and $e_0 \in [0, 0.85]$, starting with an orbit-averaged frequency $f_0 = 10$\,Hz. Dashed vertical lines mark the median mismatch, while dotted vertical lines mark the median mismatch obtained in \PaperI.
	\label{fig:Model_Mismatch_Histograms}
	}
\end{figure}

As in \PaperI, we compare the NIT dynamics at each postadiabatic order against the EOB dynamics for $M = 60 M_\odot$ on a grid of
\begin{itemize}
	\item $q \in [1, 96]$ with $\Delta q = 5$,
	\item $e_0 \in [0, 0.85]$ with $\Delta e_0 = 0.05$,
\end{itemize}
starting at periastron with an orbit-averaged frequency of $f_0 = 10$\,Hz. In addition to the 0PA, 1PA and 2PA NIT waveforms we include the OOE model of Sec.~\ref{section:OOEModel}, which integrates the oscillatory osculating-element equations with the interpolated Fourier coefficients and therefore measures the error of the interpolated data alone, without any postadiabatic truncation. Of the $360$ systems, $63$ start inside the handover region (all with $e_0 \geq 0.55$) and are excluded, as in \PaperI.

The mismatches are shown as contours in Fig.~\ref{fig:Model_Mismatch_Contours} and as histograms in Fig.~\ref{fig:Model_Mismatch_Histograms}. 
The median mismatches are $4.8 \times 10^{-5}$ (0PA), $2.1 \times 10^{-5}$ (1PA) and $9.3 \times 10^{-9}$ (2PA), with maxima of $7.9 \times 10^{-3}$, $7.0 \times 10^{-3}$ and $5.4 \times 10^{-6}$. The gain from 0PA to 1PA is modest at this total mass, whereas the 2PA model improves on both by more than three orders of magnitude in the median.

As in \PaperI, the 0PA and 1PA mismatches fare well for long inspirals with low eccentricities and large mass ratios, but become significantly worse for short inspirals with high eccentricities and comparable mass ratios, where the postadiabatic expansion converges more slowly. The 1PA mismatches agree closely with those of \PaperI\ for the same systems. The 0PA mismatches, however, are considerably smaller than in \PaperI, where their median was $2 \times 10^{-3}$, and the improvement grows with the eccentricity. This is because the two parametrizations lead to different adiabatic truncations. In \PaperI, the orbit-averaged precession of the periastron driven by radiation reaction, $\mr \avg{f_w}$, enters the evolution of the relativistic anomaly as a correction to $d\zeta/d\chi = 1$, and was therefore only included from 1PA onwards. In the $(\alpha,\beta)$ parametrization, the same term is part of $\avg{F_\alpha}$ and $\avg{F_\beta}$ and is retained already at 0PA, so that the position of the periastron is tracked at leading order. The 1PA model, in turn, improves on the 0PA model only modestly in the median, and for the most comparable masses, $q \lesssim 6$, it is in fact slightly worse. For such systems the expansion parameter is not so small, so successive orders need not improve monotonically, and the 2PA terms are needed to bring the error down to the interpolation floor. Since mismatches above $10^{-3}$ may bias parameter estimation, this motivates the use of 2PA waveforms.

The 2PA mismatch is below $10^{-6}$ for all but nine systems, all of which have $e_0 \geq 0.6$, where the signal is short and the late inspiral after the handover contributes most of the mismatch. If we examine only the inspiral portion of the waveform, the 2PA mismatch has median $4.3 \times 10^{-9}$ and maximum $3.0 \times 10^{-7}$. 

The OOE model has a median mismatch of $5.5 \times 10^{-9}$ and a maximum of $3.3 \times 10^{-7}$, and lies below the 2PA NIT for $169$ of the $297$ systems. The two agree to within a factor of a few across the whole parameter space, including the circular limit. This indicates that the 2PA averaged dynamics have practically reached the floor set by the interpolation of the tables, and a further postadiabatic order would bring no significant improvement with the present grid. Compared with \PaperI, where the 2PA waveforms at this mass reached a median of $\sim 5 \times 10^{-6}$ and a maximum of $5 \times 10^{-5}$, the median mismatch has improved by almost three orders of magnitude. This is almost entirely due to the mode tables evaluated at the NIT variables (Sec.~\ref{section:WaveformAssembly}), which remove the $\sim 10^{-6}$--$10^{-7}$ floor of evaluating the modes at the averaged variables.

\subsection{Second postadiabatic order across parameter space} \label{section:2PAOrders}
\begin{figure*}[tbp]
\includegraphics[width=0.9\textwidth]{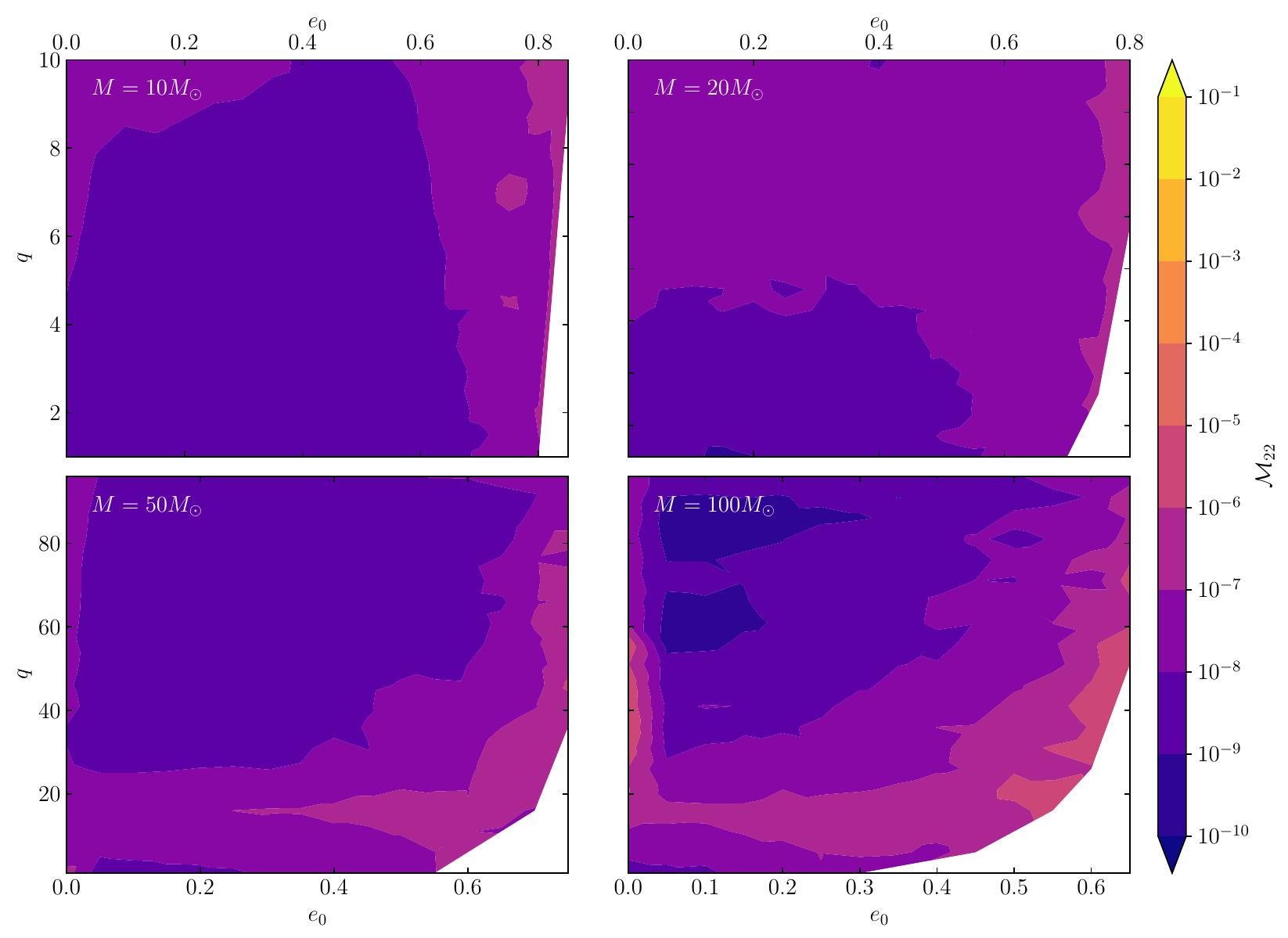}
\caption{Contour plot of the mismatch between the EOB and 2PA NIT waveforms for four total masses, over $e_0 \in [0, 0.85]$ and $q \in [1,96]$ for $M = 50$ and $100 M_\odot$, $q \in [1, 20]$ for $M = 20 M_\odot$ and $q \in [1, 10]$ for $M = 10 M_\odot$, starting with an orbit-averaged frequency $f_0 = 10$\,Hz. Mismatches use the O5 design sensitivity curve and a low-frequency cutoff of 10\,Hz. White regions are systems that begin after the NIT handover conditions are already satisfied.
\label{fig:2PA_Mismatch_Contours}
}
\end{figure*}
\begin{figure}[tbp]
\includegraphics[width=1.0\columnwidth]{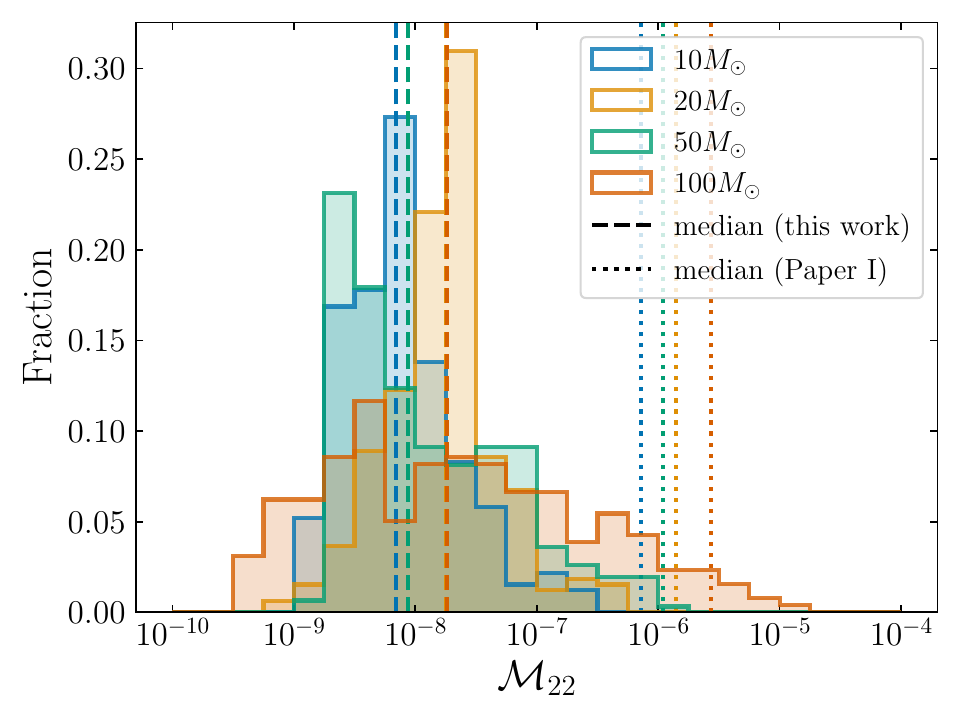}
\caption{Histogram of the mismatch between the EOB and 2PA NIT waveforms for four total masses, for the parameters of Fig.~\ref{fig:2PA_Mismatch_Contours}. Dashed vertical lines mark the median mismatch for each total mass, while dotted vertical lines mark the median mismatch obtained in \PaperI.
\label{fig:2PA_Mismatch_Histograms}
}
\end{figure}

We next examine the 2PA NIT across the parameter space of \PaperI. For each fixed total mass, we start at periastron with an orbit-averaged frequency of $f_0 = 10$\,Hz and sample the initial eccentricity $e_0 \in [0, 0.85]$ in steps of $0.05$, and the mass ratio as
\begin{itemize}
	\item $M = 100 M_\odot$, $q \in [1,96]$, $\Delta q = 5$,
	\item $M = 50 M_\odot$, $q \in [1,96]$, $\Delta q = 5$,
	\item $M = 20 M_\odot$, $q \in [1,20]$, $\Delta q = 1$,
	\item $M = 10 M_\odot$, $q \in [1,10]$, $\Delta q = 0.5$,
\end{itemize}
where the lower masses, whose signals are longest, are sampled more densely in $q$. The mismatches are shown in Figs.~\ref{fig:2PA_Mismatch_Contours} and \ref{fig:2PA_Mismatch_Histograms}. The median (maximum) mismatches are $6.9 \times 10^{-9}$ ($3.1 \times 10^{-7}$) at $M = 10 M_\odot$, $1.8 \times 10^{-8}$ ($4.0 \times 10^{-7}$) at $20 M_\odot$, $8.7 \times 10^{-9}$ ($1.5 \times 10^{-6}$) at $50 M_\odot$ and $1.8 \times 10^{-8}$ ($1.1 \times 10^{-5}$) at $100 M_\odot$. All systems with $M \leq 20 M_\odot$ have mismatches below $10^{-6}$, and the largest values at the higher masses again occur for the shortest, most eccentric signals at large mass ratios. These are the most challenging configurations for the transformations, with the highest velocities and largest eccentricities at the handover, and the few orbits that remain are then integrated by the EOB dynamics from a slightly inaccurate handover state, so that the late inspiral contributes most of the mismatch.

In contrast to \PaperI, we can seamlessly include small eccentricities and quasicircular inspirals without having to hand over to the EOB dynamics. The $e_0 = 0$ column is among the most accurate at every mass, as expected from the regularity of the $(\alpha,\beta)$ parametrization in the circular-orbit limit. The one exception is a band of mismatches between $10^{-6}$ and $10^{-5}$ at $M = 100 M_\odot$ and $e_0 = 0$ for $16 \leq q \leq 56$, visible in Fig.~\ref{fig:2PA_Mismatch_Contours}. These are the systems that reach the strong-field edge of the interpolation grid, where the handover is triggered by $\nit{p} < 8.5$ rather than by the adiabaticity condition, while evolving exactly along the $e = 0$ row of the tables. The feature is confined to $e_0 \lesssim 10^{-3}$: repeating the $q = 36$ and $q = 41$ cases at $e_0 = 0.005$ lowers the mismatch to $4$--$6 \times 10^{-9}$, in line with the rest of the column. We attribute it to the extrapolated $e = 0$ row of the reduced coefficients (Sec.~\ref{section:InterpolationGrid}) at the smallest values of $p$, and it could be removed by adding numerical nodes below $e = 0.0025$.

\begin{figure}[tbp]
\includegraphics[width=0.95\columnwidth]{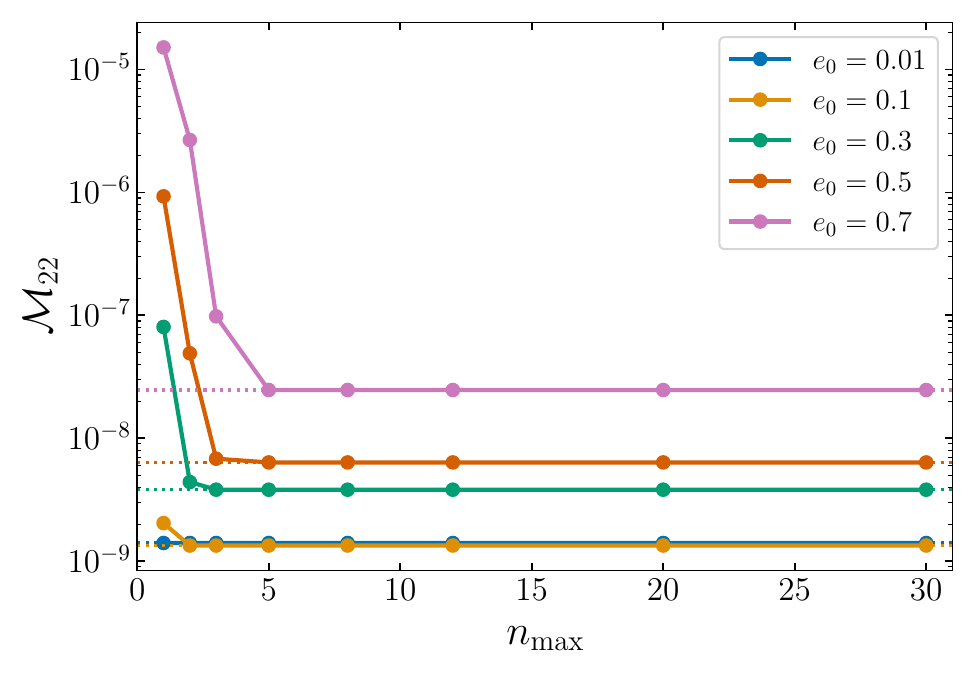}
\caption{Mismatch between the EOB and 2PA NIT waveforms as a function of the number of radial harmonics $n_\text{max}$ retained in the mode sum \eqref{eq:WaveformFourier_QC}, for $M = 20 M_\odot$, $q = 1$ and initial eccentricities $e_0 = 0.01$, $0.1$, $0.3$, $0.5$ and $0.7$. Dotted lines mark the mismatch with all $\pm 30$ interpolated harmonics.
\label{fig:HarmonicStudy}
}
\end{figure}
Finally, we examine how many radial harmonics of the mode decomposition \eqref{eq:WaveformFourier_QC} must be retained when assembling the waveform. The cost of the assembly scales with the number of harmonics evaluated per sample, so it is important to know how quickly the sum converges. Figure~\ref{fig:HarmonicStudy} shows the mismatch between the EOB and 2PA NIT waveforms for $M = 20 M_\odot$ and $q = 1$ as a function of the largest harmonic $n_\text{max}$ retained in the sum, for initial eccentricities from $e_0 = 0.01$ to $0.7$. The dotted lines mark the mismatch obtained with all $\pm 30$ interpolated harmonics, which is set by the accuracy of the 2PA dynamics rather than by the mode sum. For each eccentricity the mismatch drops steeply with $n_\text{max}$ until it reaches this floor, beyond which additional harmonics bring no further improvement. The number of harmonics needed grows with the eccentricity, as expected since the harmonic content of the modes is set by the eccentricity along the inspiral. For $e_0 \leq 0.1$ the circular term and a single harmonic already reproduce the full result ($1.3$--$1.4 \times 10^{-9}$). For $e_0 = 0.3$ one needs $n_\text{max} = 3$, and for $e_0 = 0.5$ and $0.7$ one needs $n_\text{max} = 5$, where the mismatch saturates at $6.4 \times 10^{-9}$ and $2.5 \times 10^{-8}$.

Since the eccentricity decays over the inspiral, fewer harmonics are needed as the binary approaches merger. This is exploited by the adaptive truncation of Sec.~\ref{section:WaveformAssembly}, which keeps at each point of the inspiral only the harmonics whose amplitude exceeds $10^{-6}$ of the leading one, and thereby recovers the accuracy of the full sum at a fraction of its cost. In none of the sweeps of this section did the adaptive truncation retain harmonics beyond $|n| = 8$, so mode tables with $|n| \leq 10$ would have sufficed for $e_0 \leq 0.85$. The $\pm 30$ harmonics tabulated here were chosen only because the required number was not known a priori. The small number of harmonics is a consequence of expanding in the relativistic anomaly rather than the mean anomaly, as discussed in Sec.~\ref{section:WaveformFourierDecomposition}, and is what allows the assembly to remain cheap at high eccentricity.

\subsection{Speed-up of waveform generation via interpolated Fourier decomposition} \label{section:WaveformSpeedup}
\begin{figure}[tbp]
\includegraphics[width=0.9\columnwidth]{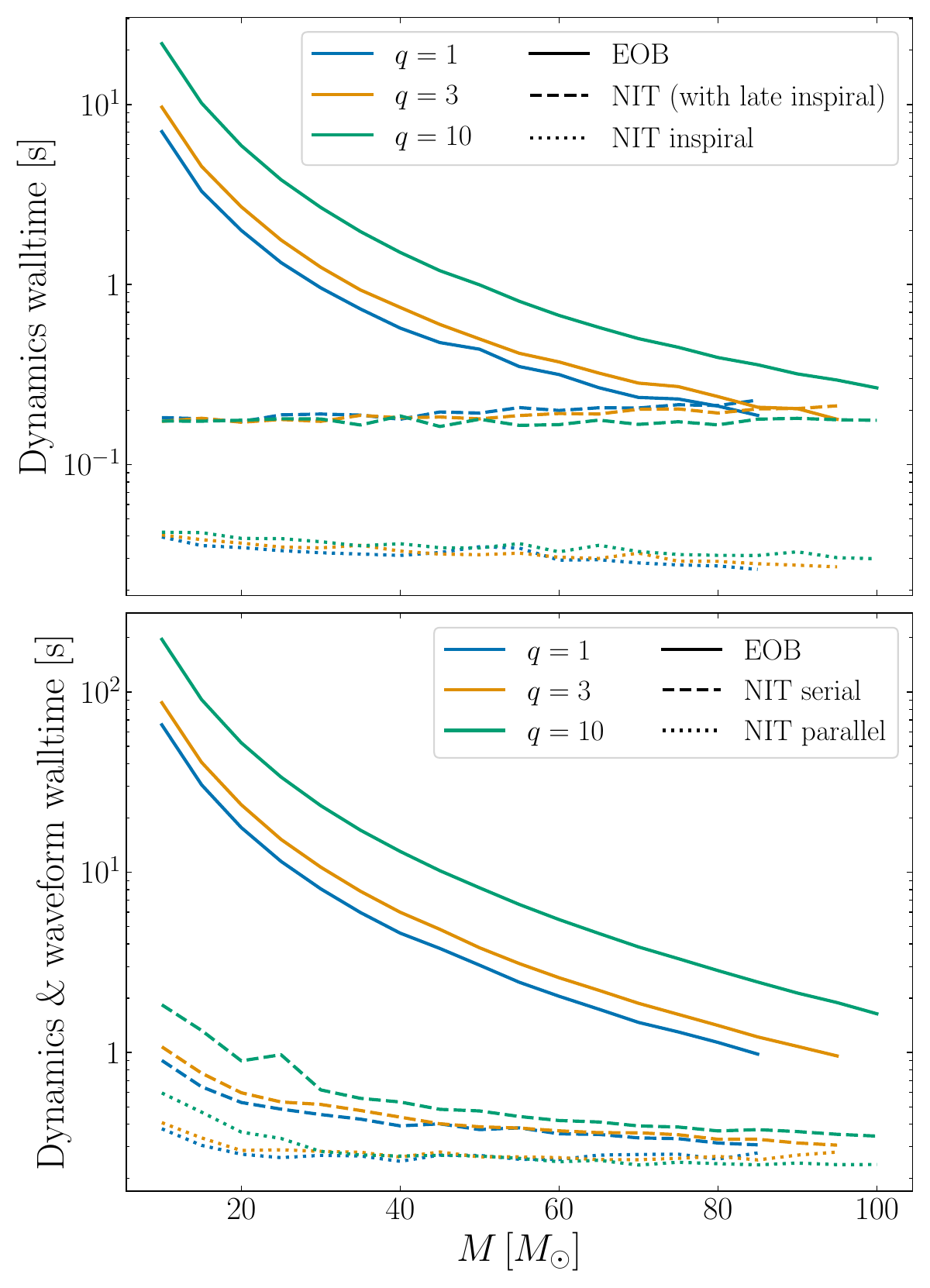}
\caption{Walltime of the dynamics (top) and of the full waveform, i.e., dynamics and waveform generation (bottom), versus total mass for systems starting at periastron with an orbit-averaged initial frequency of $10$\,Hz and initial eccentricity $e_0 = 0.4$, for $q = 1$, $3$ and $10$. Top: EOB dynamics (solid), NIT dynamics including the EOB late inspiral (dashed) and NIT inspiral alone (dotted). Bottom: EOB waveform (solid) and 2PA NIT waveform assembled serially (dashed) and on eight parallel kernels (dotted).}
\label{fig:Walltime_vs_Mass}
\end{figure}
\begin{figure}[tbp]
\includegraphics[width=0.9\columnwidth]{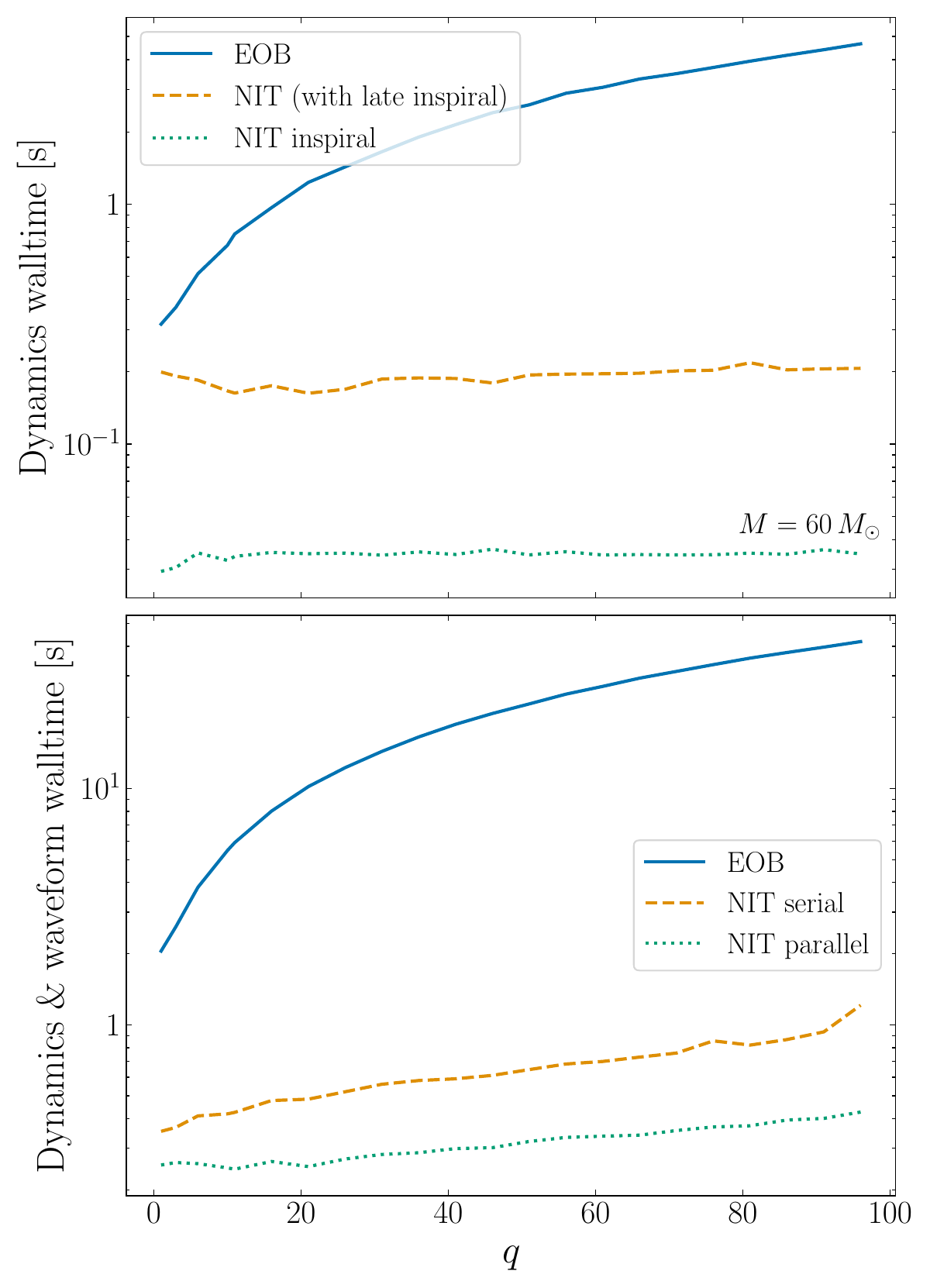}
\caption{Walltime of the dynamics (top) and of the full waveform, i.e., dynamics and waveform generation (bottom), versus mass ratio for $M = 60 M_\odot$, starting at periastron with an orbit-averaged initial frequency of $10$\,Hz and $e_0 = 0.4$. Top: EOB dynamics (solid), NIT dynamics including the EOB late inspiral (dashed) and NIT inspiral alone (dotted). Bottom: EOB waveform (solid) and 2PA NIT waveform assembled serially (dashed) and on eight parallel kernels (dotted).}
\label{fig:Walltime_vs_MassRatio}
\end{figure}
\begin{figure}[tbp]
\includegraphics[width=0.95\columnwidth]{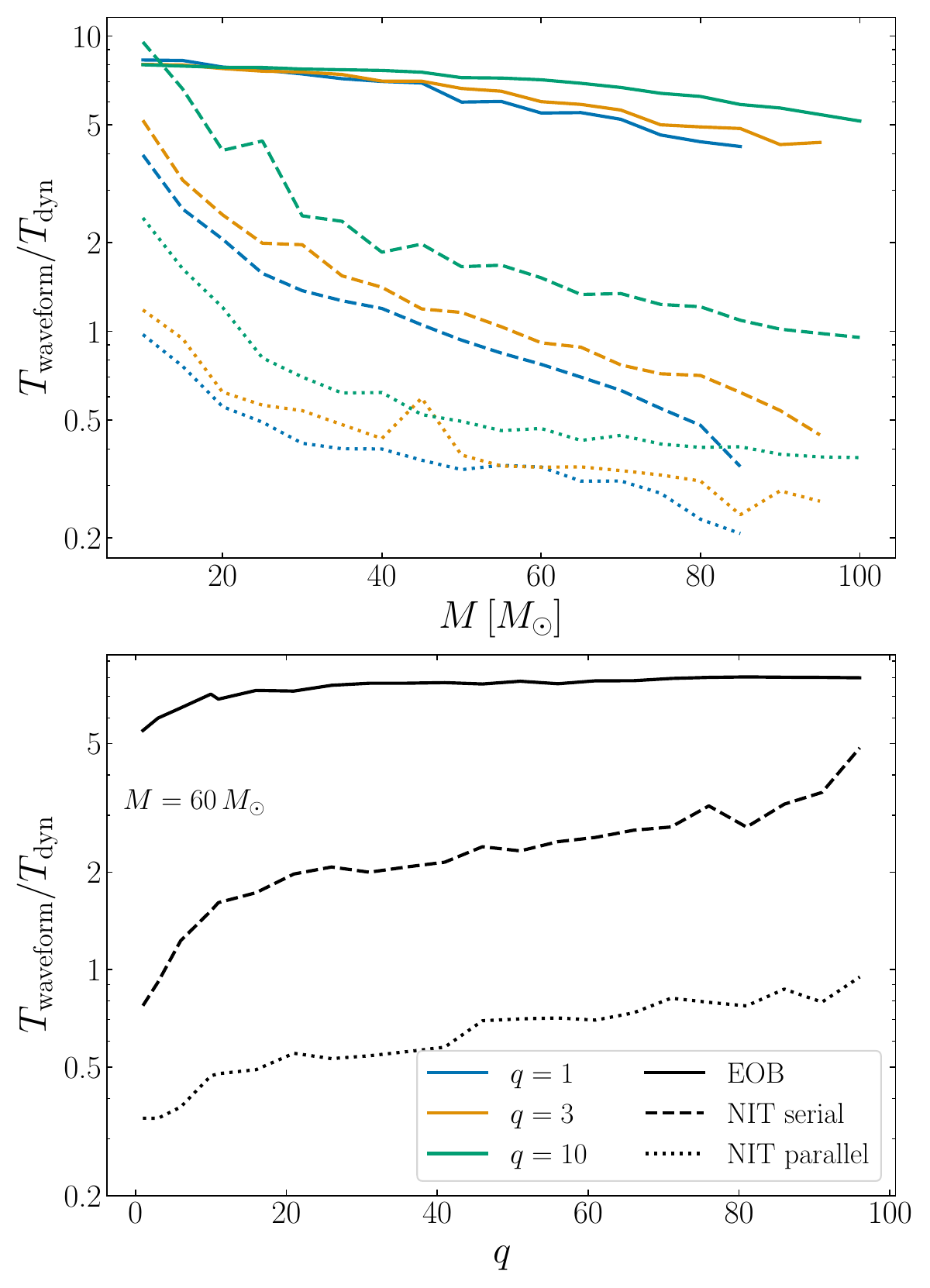}
\caption{Cost of the waveform generation relative to that of the dynamics, $T_\text{waveform}/T_\text{dyn} = T_\text{total}/T_\text{dyn} - 1$, for the systems of Figs.~\ref{fig:Walltime_vs_Mass} and \ref{fig:Walltime_vs_MassRatio}: versus total mass for $q = 1$, $3$ and $10$ (top) and versus mass ratio for $M = 60 M_\odot$ (bottom), for the EOB waveform (solid) and the 2PA NIT waveform assembled serially (dashed) and on eight parallel kernels (dotted).}
\label{fig:Waveform_to_Dynamics_Ratio}
\end{figure}
\begin{figure}[tbp]
\includegraphics[width=0.95\columnwidth]{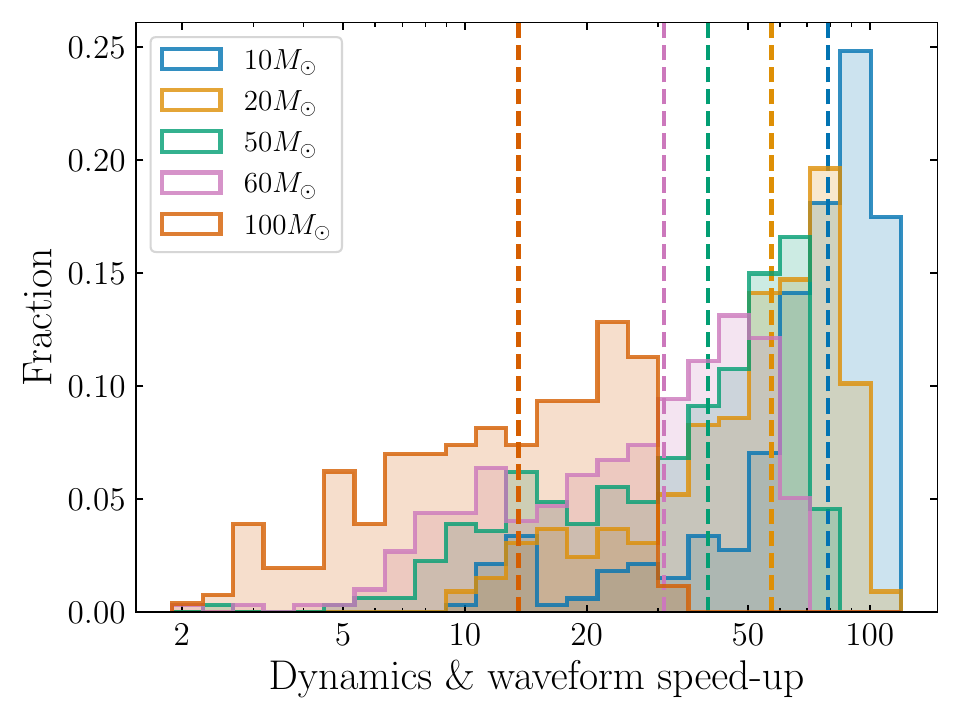}
\caption{Total speed-up of the dynamics and waveform generation from using the 2PA NIT rather than the EOB dynamics for the systems of Sec.~\ref{section:2PAOrders}, recorded during the mismatch sweeps. The dashed vertical lines mark the median speed-up for each total mass.
\label{fig:Total_Speedup_Histograms}
}
\end{figure}
Finally, we turn to the computational cost. As in \PaperI, all timings are recorded on an Apple M1 Max.

The top panel of Fig.~\ref{fig:Walltime_vs_Mass} shows the walltime of the dynamics alone as a function of the total mass, for systems with $e_0 = 0.4$ starting at $10$\,Hz with $q = 1$, $3$ and $10$ and total masses between $10$ and $100 M_\odot$. The curves for $q = 1$ and $q = 3$ end at $M = 85$ and $95 M_\odot$ respectively, beyond which the systems start inside the handover region. The cost of the EOB dynamics is proportional to the number of orbits in band, and so falls steeply with the total mass, from $7$--$22$\,s at $M = 10 M_\odot$ to $0.2$--$0.3$\,s at the highest masses, and grows with the mass ratio at fixed mass. The NIT dynamics, by contrast, are integrated on the radiation-reaction timescale and take $0.16$--$0.23$\,s regardless of the length of the signal. Most of this is the EOB integration of the late inspiral after the handover. The averaged inspiral itself takes only $0.03$--$0.04$\,s, as shown by the dotted lines, and is the only part of the dynamics that is not already present in the EOB model. The speed-up of the dynamics therefore ranges from $125\times$ for the longest signal ($M = 10 M_\odot$, $q = 10$) down to order unity at $M \gtrsim 80 M_\odot$, where the EOB inspiral is itself only a few hundred milliseconds long and the curves cross.

The bottom panel of Fig.~\ref{fig:Walltime_vs_Mass} shows the total walltime of the full waveform, i.e., the dynamics, the waveform modes and the late inspiral, for the same systems. Each configuration is timed on a single core, and the 2PA waveform additionally with the assembly distributed over eight kernels. For the EOB model the mode expression is evaluated at every step of the dynamics, and this dominates the cost by roughly an order of magnitude, giving walltimes from $196$\,s at $M = 10 M_\odot$ and $q = 10$ down to $1$\,s at the highest masses. The number of integrator steps is comparable to the number of output samples, so this cost is set by the price of a single evaluation of the symbolic 3PN mode expression in Mathematica, about $180\,\mu$s. We retain the waveform generation method of \PaperI\ for continuity with that work, so the end-to-end speed-ups quoted here would be optimistic upper bounds if applied to an optimized production model, while the speed-up of the dynamics would be the number that carries over. The 2PA waveform takes $0.3$--$1.8$\,s serially and $0.24$--$0.6$\,s parallelized on eight kernels. Its dependence on the total mass is much weaker, since the fixed cost of about $0.2$--$0.25$\,s (Sec.~\ref{section:WaveformSettings}) dominates for short signals and only the assembly on the time grid scales with the number of samples. Parallelizing the assembly therefore helps most for the longest signals, reducing the walltime at $M = 10 M_\odot$ and $q = 10$ from $1.8$ to $0.6$\,s, and hardly at all above $M \sim 60 M_\odot$, where the fixed cost dominates. The resulting total speed-up is $3$--$100\times$ serially and up to $300\times$ for the parallel assembly against the serial EOB waveform, increasing with the number of orbits in band, i.e., with decreasing total mass and increasing mass ratio.

Figure~\ref{fig:Walltime_vs_MassRatio} shows the same quantities as a function of the mass ratio, for $M = 60 M_\odot$, $e_0 = 0.4$ and $q$ between $1$ and $96$. The number of orbits in band grows roughly as $1/\nu$, and so the walltime of the EOB dynamics rises from $0.32$\,s at $q = 1$ to $4.7$\,s at $q = 96$, and that of the EOB waveform from $2.1$ to $42$\,s. The NIT dynamics remain flat at $0.16$--$0.22$\,s, with the averaged inspiral again contributing only $0.03$--$0.04$\,s, so that the speed-up of the dynamics grows from $1.6\times$ at $q = 1$ to $23\times$ at $q = 96$. The total walltime of the 2PA waveform grows only mildly with the mass ratio, from $0.35$ to $1.2$\,s serially and from $0.25$ to $0.43$\,s in parallel, as the longer signals require more samples to be assembled. The total speed-up thus increases from $6\times$ at $q = 1$ to about $40\times$ for $q \gtrsim 60$ serially, and from $8\times$ to $99\times$ in parallel.

Figure~\ref{fig:Waveform_to_Dynamics_Ratio} shows how much of the total cost is spent on generating the waveform rather than the dynamics, for the same systems. For the EOB model the waveform costs $4$--$8$ times the dynamics for every system, since the mode expression is evaluated at every step. For the 2PA NIT, the ratio spans a much wider range. It is below one for short signals at high masses, where the fixed cost of the late inspiral dominates, but rises with the number of orbits in band to about $4$ at $M = 10 M_\odot$ for $q = 1$ and almost $10$ for $q = 10$ serially, and to $1$--$2.4$ in parallel. The waveform assembly is therefore still the dominant cost for the longest signals, even though it is now cheap per sample, and any further gain for such systems must come from the assembly, for example through the GPU parallelization discussed in Sec.~\ref{section:Discussion}.

Figure~\ref{fig:Total_Speedup_Histograms} shows the total speed-up of the waveform generation recorded for every system of the mismatch sweeps of Sec.~\ref{section:2PAOrders}, which cover the full range of eccentricities and mass ratios at $M = 10$, $20$, $50$, $60$ and $100 M_\odot$. Here both NIT and EOB waveform generators use the parallel assembly, and each system was timed once on a machine that was not otherwise idle, so the distributions are indicative rather than benchmarks. The medians are $79\times$, $57\times$, $40\times$, $31\times$ and $14\times$ for the five masses, with the 2PA waveforms taking a median of $0.5$--$1.2$\,s against $6.5$--$93$\,s for the EOB waveforms. The distributions shift towards lower speed-ups as the total mass increases and the number of orbits in band decreases. They are skewed towards their upper ends, which are set by the systems with moderate eccentricities, $0.1 \lesssim e_0 \lesssim 0.4$, and large mass ratios, and the largest speed-up at each mass lies between $32\times$ at $M = 100 M_\odot$ and $120\times$ at $M = 10 M_\odot$. The low tails come from two groups of systems. The first are the circular inspirals with $e_0 = 0$, for which the EOB dynamics have no periastron passages to resolve and are several times cheaper than for a slightly eccentric system of the same mass and mass ratio, while the cost of the 2PA waveform is set by the length of the signal. The second are the most eccentric systems, whose signals are short enough that the fixed cost of the 2PA waveform dominates. For both groups the speed-up falls to $2$--$13\times$ depending on the total mass.

In summary, the 2PA NIT waveform takes between $0.25$ and $2$\,s for any system considered here, compared to up to several minutes for the EOB waveform of the same system, and the speed-up grows with the number of orbits in band. This is to be compared with the $1.5$--$8\times$ of \PaperI, where the evaluation of the mode expressions on a dense $\chi$ grid limited the gain. With the interpolated mode decomposition and the direct assembly on the time grid, the speed-up of the dynamics is largely carried over to the full waveform, although for the longest signals the assembly of the waveform still costs several times more than the averaged dynamics (Fig.~\ref{fig:Waveform_to_Dynamics_Ratio}). What remains is a fixed cost of about $0.2$--$0.25$\,s per waveform, dominated by the EOB integration of the late inspiral, which sets the floor for short signals, and the assembly cost of about $1.6$\,$\mu$s per sample, which can be shared across cores for long signals. Independently of the reference, this assembly cost is about a hundred times lower than the roughly $200\,\mu$s per point of the interpolant-based generator of \PaperI\ in the same environment, and the $10^{-6}$--$10^{-7}$ mismatch floor of that generator is gone.

\section{Discussion} \label{section:Discussion}
In this paper, we have extended the near-identity averaging framework of \PaperI\ to the small-eccentricity and quasicircular limits. The parametrization of \PaperI\ in terms of the eccentricity and the relativistic anomaly becomes singular as $e \rightarrow 0$, where the argument of periapsis is ill-defined and the osculating eccentricity ceases to be a small oscillation about its average, which forced \PaperI\ to hand over to the full EOB dynamics at $\tilde{e} < 0.035$. By replacing $(e,w)$ with the components of the eccentricity vector $(\alpha,\beta) = e (\sin w, \cos w)$, we obtained osculating-element equations of motion and near-identity transformations through second postadiabatic order that are regular through the circular-orbit limit, where, as shown in Appendix~\ref{appendix:OOE_EoM_QC}, $\chi$ becomes the phase of the radial epicyclic oscillation and the eccentricity vector is driven by radiation reaction at the finite rate $\dot{p}/p$. The rotational covariance of the equations under a shift of the periastron reduces the resulting four-dimensional parameter space back to the three-dimensional grid of \PaperI, so that all averaged terms and transformation functions could be built from the existing Fourier coefficients, supplemented by high-precision data at small eccentricities. We further introduced a new waveform-generation method in which the waveform modes are decomposed into radial harmonics tabulated directly as functions of the averaged variables, and the waveform is assembled on the output time grid by inverting the time transformation sample by sample, removing both the evaluation of the mode expressions along the inspiral and the $\sim 10^{-6}$ mismatch floor incurred in \PaperI\ by evaluating the modes at the averaged variables.

Comparing against waveforms generated from the full EOB evolution of the same inspiral-only, nonspinning model over the parameter space of \PaperI, extended to $e_0 = 0$, the 2PA NIT waveforms reach median mismatches of $\sim 10^{-8}$ and maximum mismatches of $\sim 10^{-5}$ for total masses between $10$ and $100 M_\odot$, including quasicircular inspirals, which no longer require a handover to the EOB dynamics at small eccentricities; the remaining error is set by the interpolation of the tables rather than by the postadiabatic truncation, as the comparison with the OOE model shows. At the same time, the full waveform is generated $3$--$100$ times faster than with the EOB dynamics on a single core, and up to $300$ times faster with the assembly distributed over eight cores against the serial EOB waveform. With both generators parallelized, the sweeps give median speed-ups of $14$--$79\times$ and maxima of up to $\sim 120\times$. The gain grows with the number of orbits in band, so the method is most effective precisely for the low-mass, large-mass-ratio systems that are most expensive to model. The fixed cost of about a quarter of a second per waveform, spent on the averaged dynamics, the evaluation of the Fourier coefficients and the EOB late inspiral, sets the computational floor for short signals. All timings refer to our proof-of-concept implementation in Mathematica, and the end-to-end speed-ups depend strongly on how optimized the reference waveform generation is. Our reference evaluates the symbolic mode expression at every integrator step, which makes its waveform cost several times its dynamics (Sec.~\ref{section:WaveformSpeedup}), whereas in production codes, implemented in Python, C++, or other compiled languages, the dynamics dominate and the waveform costs a fraction of them. For such a code the gain to expect is closer to the speed-up of the dynamics weighted by the fraction of the walltime spent on the dynamics than to the end-to-end numbers quoted here. The improvements over \PaperI\ that do not depend on the reference are the absolute walltime of the NIT waveform, the assembly cost per sample, which is about a hundred times lower than in \PaperI, and the removal of its $10^{-6}$--$10^{-7}$ mismatch floor.

Several improvements remain possible. With the inspiral now cheap, the EOB evolution of the late inspiral after the handover accounts for a sizeable fraction of the remaining cost, and the largest mismatches (all at high eccentricity and mass ratio, where the handover occurs in the most relativistic configurations) arise from the same segment. The few relatively high mismatches at $M = 100 M_\odot$ and $e_0 = 0$ are a result of our particular implementation choices and can be remedied with more numerical information. The handover conditions of Sec.~\ref{section:Handover}, inherited from \PaperI, are conservative and could be relaxed to allow a longer NIT evolution and a shorter EOB one. A more physically motivated treatment of the eccentric transition-to-plunge \cite{Becker:2024xdi,Lhost:2024jmw} could further improve both the accuracy and the speed of the resulting waveforms. Moreover, we have modeled only the $(2,2)$ mode. The Fourier decomposition and the assembly on the time grid apply unchanged to the higher modes, each of which adds its own set of coefficient tables and a proportional cost per sample, and we leave their inclusion to future work.

The $(\alpha,\beta)$ parametrization is as valid for large eccentricities as it is for small eccentricities. As in \PaperI, only eccentricities $e \leq 0.9$ were considered, this being the extent of the interpolation grid, and most systems with $e_0 \gtrsim 0.7$ at $10$\,Hz start inside the handover margin and were excluded. Methods that rely on orbit averaging are ill-suited to very high eccentricities \cite{McCart:2021upc,Fumagalli:2025rhc}, and a Fourier decomposition with respect to the relativistic anomaly still scales poorly with eccentricity. A dedicated study that does not rely on the Fourier decomposition and interpolates to much higher eccentricities would be needed to establish the exact limits of NITs in that regime.

While we have reduced the cost of waveform generation by orders of magnitude for low masses and large mass ratios, the longest signals, those of extreme mass ratios, remain dominated by the waveform assembly, whose cost scales with the number of samples. We have shown that this can be reduced via parallelization on multiple CPU cores. One can go further and follow the lead of the \texttt{FastEMRIWaveforms} package \cite{Chua:2020stf,Katz:2021yft,Speri:2023jte,Chapman-Bird:2025xtd} and leverage multithreading on GPUs. As outlined in Sec.~\ref{section:WaveformAssembly}, waveform assembly is now a fixed, small amount of arithmetic per sample with no dependence between samples, so it can be distributed across many CPU or GPU threads with only the sparse coefficient data shared between them. Moreover, our use of the relativistic anomaly rather than the mean anomaly as the radial phase dramatically reduces the number of harmonics needed to reconstruct the waveform (five suffice up to $e_0 = 0.7$, Sec.~\ref{section:2PAOrders}), so that the approach may scale even more favorably with eccentricity than the multi-voice decomposition in the mean anomaly. 

Complementary approaches that reduce the need for dense time-domain sampling, such as nonuniform sampling, frequency-domain waveforms via the stationary phase approximation \cite{Haberland:2026xvj} or shifted uniform asymptotics \cite{Morras:2025nlp,Klein:2014bua}, or time-frequency representations \cite{Cornish:2020odn,Bandopadhyay:2025fyx,Cornish:2026mrt}, or batching several waveform calls in parallel on a GPU, could equally be applied together with the averaged dynamics and waveform assembly presented here.

The averaged dynamics of this work connect naturally to the eccentric dynamics of EMRIs modeled with the gravitational self-force formalism. The regularity of the $(\alpha,\beta)$ variables in the circular limit means that adiabatic self-force information can be incorporated across the full range of eccentricities, including quasicircular inspirals, while our NIT expansion supplies the postadiabatic corrections beyond the 1PA order currently used in EMRI modeling. Many efforts have already used information from the test-mass limit to improve EOB waveforms at large mass ratios \cite{Albertini:2022dmc,Albertini:2023aol,Albertini:2024rrs,Albertini:2024agg,Faggioli:2025hff,Nishimura:2026nse,Faggioli:2026alx}. Our framework offers new calibration methods such as calibrating the test-mass limit of the eccentric EOB radiation-reaction force to eccentric self-force flux calculations \cite{Faggioli:2024ugn}. In the test-mass limit, the averaged forcing functions of the orbital elements are precisely the orbit-averaged fluxes of energy and angular momentum computed in self-force theory, so that the comparison, and any calibration, can be made directly in these variables, whereas the instantaneous radiation-reaction force cannot be compared with adiabatic fluxes point by point. This would reduce the dominant error of EOB waveforms for extreme mass ratios, which stems from missing adiabatic-order information \cite{Khalvati:2025znb}. Conversely, adiabatic self-force models could be supplemented with postadiabatic information from eccentric EOB models to produce hybrid models for the intermediate-mass-ratio regime.

This framework of osculating orbital elements and near-identity transformations is agnostic to the specific EOB model, as long as the equations of motion take the form of Eqs.~\eqref{eq:EOB_EOM}. Having successfully demonstrated it on an inspiral-only EOB model, we plan to apply this framework to state-of-the-art spinning SEOBNR waveform models to accelerate both the dynamics and the waveform generation. The techniques in this paper extend naturally to the aligned-spin model \texttt{SEOBNRv6EHM} \cite{Gamboa:2026jht}. The only difficulty is the interpolation over the two additional spin parameters, for which efficient multidimensional interpolation methods are well established \cite{christian_chapman_bird_2025_15231132,Khalvati:2025znb}. For the precessing systems of \texttt{SEOBNRv6EPHM} \cite{Gamboa:2026ybr}, we aim to account for the coupling between the eccentric orbital dynamics and the spin precession. The orientation of the orbital plane then becomes an additional slowly varying element, the precession frequency introduces a third timescale between the orbital and radiation-reaction ones, and the spin evolution equations must be averaged alongside the orbital elements. This may require us to average over multiple timescales \cite{Gerosa:2023xsx,Drummond:2023wqc} and treat transient orbital resonances, as has been done in the context of EMRIs \cite{Lynch:2024ohd}.

Overall, the technique presented here has the potential to reduce the cost of generating eccentric EOB waveforms by orders of magnitude for low-mass and large-mass-ratio systems, all the way to the circular limit, without sacrificing the accuracy required for data analysis. This will be particularly valuable for future gravitational-wave detectors, which will observe many more low-mass and large-mass-ratio systems, over many more orbits, than current detectors.

During the final stages of this work, we became aware of Ref.~\cite{Gamba:2026ado}, where the author adopts equinoctial elements, regular in the circular limit, and extends the osculating orbit scheme of Ref.~\cite{Damour:2004bz} with near-identity transformations to next-to-next-to-leading order, equivalent to our 2PA. This framework is then applied to the \texttt{TEOBResumS-Dal\'{\i}} model \cite{Albanesi:2025txj}, showing that the method speeds up the dynamics of a production-level spinning EOB model with tidal effects while retaining mismatches $< 5 \times 10^{-7}$. That work does not tabulate the averaged equations of motion offline and does not employ a fast waveform-generation scheme. Its speed-ups are measured against a different baseline, a highly optimized C implementation of a full inspiral-merger-ringdown model with spins and tides, and are therefore not directly comparable with ours. Against that baseline the median speed-ups are $\sim 1.5$ for BBHs and $\sim 2.6$ for binary neutron stars, with a median binary-neutron-star waveform taking $\sim 0.5$\,s.

\begin{acknowledgments}
PL thanks C. Kavanagh for helpful discussions at the early stages of the project and R. Gamba for their comments on the manuscript. 
Funded by the European Union (ERC grant GWSky/ 101167314). Views and opinions expressed are however those of the author(s) only and do not necessarily reflect those of the European Union or the European Research Council Executive Agency. Neither the European Union nor the granting authority can be held responsible for them. 
MvdM further acknowledges financial support by 
the VILLUM Foundation (grant no. VIL37766), and the DNRF Chair program (grant no. DNRF162) by the Danish National Research Foundation.
The Center of Gravity is a Center of Excellence funded by the Danish National Research Foundation under grant No. 184.
The authors acknowledge the use of \texttt{Claude} \cite{Anthropic:2026Claude} to assist in refactoring and optimizing code and in proofreading the manuscript. All code and text were reviewed by the authors, who take full responsibility for the content.
\end{acknowledgments}


\appendix
\section{Quasicircular limit of osculating orbital element equations of motion} \label{appendix:OOE_EoM_QC}
We will now examine the quasicircular limit of the osculating orbital element equations of motion \eqref{eq:OOE_EoM_QC}. Throughout this appendix we write $r_0 \equiv p M$ for the radius of the tangent circular orbit, denote by $A$, $A'$, and $A''$ the potential $A(\nu;r)$ and its first and second derivatives with respect to $r$ evaluated at $r = r_0$, $\bar{D}$ denotes the potential $\bar{D}(\nu;r)$ evaluated at $r = r_0$, and label the leading-order coefficients of an expansion in $e$ by a superscript in parentheses, e.g.\ $X = X^{(0)} + e X^{(1)} + \mathcal{O}(e^2)$. Recall that the turning points are $r_1 = r_0/(1-e)$ (apastron) and $r_2 = r_0/(1+e)$ (periastron), so that both tend to $r_0$ as $e \rightarrow 0$. 

First, we begin by expanding the expression for $H_{\rm eff}$ in powers of $e$:
\begin{align}
\begin{split}
H_{\rm eff} & =  \frac{2 \sqrt{e} \sqrt{A(r_1) A(r_2)}}{\sqrt{(1+e)^2 A(r_2) - (1 - e)^2 A(r_1)}} \\ & = H_{\rm eff}^{(0)} + \mathcal{O}(e^2) = \frac{\sqrt{2}\, A}{\sqrt{2 A - r_0 A'}} + \mathcal{O}(e^2) .
\end{split}
\end{align}
From this one can also trivially find the circular orbit limit of the EOB Hamiltonian, $H_{\rm EOB}^{(0)} = M \sqrt{1 + 2 \nu (H_{\rm eff}^{(0)} - 1) }$.
Similarly, we can expand the expression for $p_\phi$ in powers of $e$:
\begin{align}
\begin{split}
		p_\phi & = \sqrt{\frac{r_0^2 \left[A(r_2) - A(r_1)\right]}{(1-e)^2 A(r_1) - (1+e)^2 A(r_2)}} \\ & =  p_{\phi}^{(0)} + \mathcal{O}(e^2) = \frac{r_0 \sqrt{r_0 A'}}{\sqrt{2 A - r_0 A'}} + \mathcal{O}(e^2) .
\end{split}
\end{align}
Note that neither $H_{\rm eff}$ nor $p_\phi$ has a term linear in $e$.

Using the azimuthal component of the osculating condition \eqref{eq:osculating_momentum}, one can find the circular orbit limit of the rate of change of $p$ by rearranging $\dot{p}_\phi = (\partial p_\phi/\partial p)\, \dot{p} + (\partial p_\phi/\partial e)\, \dot{e} = F_\phi$; since $\partial p_\phi / \partial e = \mathcal{O}(e)$, we obtain
\begin{equation}
 \frac{d p}{dt} = \dot{p}^{(0)} + \mathcal{O}(e) =  \left(\frac{d p_\phi^{(0)}}{d p}\right)^{-1} F_\phi^{(0)} + \mathcal{O}(e),
\end{equation}
where
\begin{equation}
\dot{p}^{(0)} = \frac{\left(2 A - r_0 A'\right)^{3/2} \sqrt{r_0 A'}}{M r_0 \left[ A \left(3 A' + r_0 A''\right) - 2 r_0 A'^2 \right]}\, F_\phi^{(0)} .
\end{equation}

To find the leading-order-in-eccentricity expression for $p_{r_*}$, we first note that $p_{r_*}(e = 0) = 0$, meaning that $p_{r_*} = e\, p_{r_*}^{(1)} + \mathcal{O}(e^2)$ for small $e$. Thus if we rearrange Eq.~\eqref{eq:Heff} we can justify dropping the $Q$ term, which is $\mathcal{O}(p_{r_*}^4)$ and contributes only at higher order in $e$. This leaves us with
\begin{equation}
p_{r_*}^2 = H_{\rm eff}^{2} - A(r) \left(1 + \frac{p_{\phi}^{2}}{r^2}\right) + \mathcal{O}(e^4) .
\end{equation}
Expanding the right-hand side to $\mathcal{O}(e^2)$, with $r = r_0/(1 + e \cos \zeta)$ and including the $\mathcal{O}(e^2)$ terms of $H_{\rm eff}^2$ and $p_\phi^2$ (which are needed for $p_{r_*}$ to vanish at the turning points $\zeta = 0, \pi$), and using $p_r = p_{r_*}/\xi(r)$ with $\xi = A(\nu;r) \sqrt{\bar{D}(\nu;r)}$, we find $p_{r}  = e\, p_{r}^{(1)} \sin \zeta + \mathcal{O}(e^2)$, where
\begin{equation}
p_{r}^{(1)} = \frac{1}{A \sqrt{\bar{D}}} \sqrt{\frac{r_0 \left(r_0 A A''-2 r_0 A'^2+3 A A'\right)}{2 A - r_0 A'}} .
\end{equation}

We then use this expression to examine the leading order in the rate of change of $e$ in the quasicircular limit.
First, using the radial component of the osculating condition \eqref{eq:osculating_position}, $(\partial r/\partial p)\, \dot{p} + (\partial r / \partial e)\, \dot{e} + (\partial r/\partial w)\, \dot{w} = 0$ with $\partial r / \partial w = - \partial r / \partial \zeta$, we can write
\begin{align}
\begin{split} \label{eq:wdot_expansion}
\dot{w} & = \frac{1}{\partial r / \partial \zeta} \left( \frac{\partial r} {\partial p} \dot{p} + \frac{\partial r} {\partial e} \dot{e} \right) \\
& = \dot{w}^{(-1)} e^{-1} + \dot{w}^{(0)} + \mathcal{O}(e) \\
& = \frac{1}{e} \left( \frac{\dot{p}^{(0)}}{p \sin \zeta} - \dot{e}^{(0)} \cot \zeta \right) + \frac{\dot{p}^{(0)} \cot \zeta}{p} + \mathcal{O}\left(e\right) .
\end{split}
\end{align}
This leads to the conclusion that $\dot{w}$ diverges as $e \rightarrow 0$, which is consistent with $w$ being ill-defined in the circular orbit limit.

Next, using the radial component of the osculating condition \eqref{eq:osculating_momentum}, $(\partial p_r/\partial p)\, \dot{p} + (\partial p_r / \partial e)\, \dot{e} + (\partial p_r/\partial w)\, \dot{w} = F_r$, and keeping only the leading order in $e$, we obtain
\begin{equation}
p_r^{(1)} \sin \zeta\; \dot{e}^{(0)} - p_r^{(1)} \cos \zeta\; \dot{w}^{(-1)} = F_r^{(0)} = 0 ,
\end{equation}
where the last equality follows because every term of the radial component of the radiation-reaction force (Appendix~C of \PaperI) carries a factor of $p_r$, so that $F_r = \mathcal{O}(e)$ for nearly circular orbits. Substituting in Eq.~\eqref{eq:wdot_expansion} and solving for $\dot{e}^{(0)}$ we find
\begin{equation}
\dot{e} = \frac{\dot{p}^{(0)} \cos \zeta}{p} + \mathcal{O}(e), \qquad \dot{w} = \frac{\dot{p}^{(0)} \sin \zeta}{e\, p} + \mathcal{O}(1) .
\end{equation}
As such, in the osculating orbital element equations of motion, the rate of change of $e$ is non-zero and oscillatory in the quasicircular limit, though its orbit-averaged rate of change is zero. 

From here, one can find the leading-order-in-eccentricity expression for $\dot{\alpha}$, which is given by 
\begin{align}
\begin{split}
\dot{\alpha} & = \dot{e} \sin w + e \dot{w} \cos w \\
& = \frac{\dot{p}^{(0)}}{p} \left( \cos \zeta \sin w + \sin \zeta \cos w \right) + \mathcal{O}(e) \\
& = \frac{\dot{p}^{(0)} \sin \chi}{p} +\mathcal{O}\left(e\right) .
\end{split}
\end{align}
Similarly for $\dot{\beta}$:
\begin{align}
\begin{split}
\dot{\beta} & = \dot{e} \cos w - e \dot{w} \sin w \\
& = \frac{\dot{p}^{(0)}}{p} \left( \cos \zeta \cos w - \sin \zeta \sin w \right) + \mathcal{O}(e) \\
& = \frac{\dot{p}^{(0)} \cos \chi}{p} +\mathcal{O}\left(e\right) .
\end{split}
\end{align}
From this we can see that the rates of change of $\alpha$ and $\beta$ are both finite and non-zero oscillations in the quasicircular limit, but with zero orbit average: at leading order the eccentricity vector $(\alpha,\beta)$ is driven at the rate $\dot{p}^{(0)}/p$ in a direction that rotates with the radial phase $\chi$. This is consistent with the fact that $\alpha$ and $\beta$ are well-defined in the circular orbit limit, and it is the origin of the oscillation of the osculating eccentricity that persists as $\tilde{e} \rightarrow 0$ discussed in Sec.~\ref{section:OOE_Summary}. The radial radiation-reaction force first enters at the next order, through $F_r = e F_r^{(1)} \sin \zeta + \mathcal{O}(e^2)$.

We now move on to the extrinsic quantities, starting with the rate of change of $\phi$. With the momenta normalized by $\mu$ as in Sec.~\ref{section:EOB_Dynamics}, Hamilton's equations read $\dot{\phi} = (M/H_{\rm EOB})\, \partial H_{\rm eff}/\partial p_\phi$ and $\dot{r} = \xi(r) (M/H_{\rm EOB})\, \partial H_{\rm eff}/\partial p_{r_*}$. One can straightforwardly expand the former in powers of $e$ to find
\begin{align}
\begin{split}
\dot{\phi} & = \dot{\phi}^{(0)} + \mathcal{O}(e) = \frac{M A\, p_\phi^{(0)}}{r_0^2 H_{\rm eff}^{(0)} H_{\rm EOB}^{(0)}} + \mathcal{O}(e) \\
& = \frac{M A \sqrt{r_0 A'}}{r_0 \sqrt{2 A - r_0 A'}\, H_{\rm eff}^{(0)} H_{\rm EOB}^{(0)}} + \mathcal{O}\left(e\right) ,
\end{split}
\end{align}
which is the orbital frequency of the circular orbit at $r_0$. Similarly, expanding the rate of change of $r$ (the $Q$ term again contributing only at higher order) we find
\begin{align}
\begin{split}
\dot{r} & = \dot{r}^{(1)} e \sin \zeta + \mathcal{O}(e^2) \\
& = \frac{M A^2 \bar{D}\, p_r^{(1)}}{H_{\rm eff}^{(0)} H_{\rm EOB}^{(0)}}\, e \sin \zeta +\mathcal{O}\left(e^2\right) .
\end{split}
\end{align}

With this in hand, we can find the rate of change of the time coordinate $t$ with respect to $\chi$ in the quasicircular limit:
\begin{align}
\begin{split}
\frac{dt}{d\chi} & = \frac{\partial r / \partial \zeta}{\dot{r}} = f_t^{(0)} + \mathcal{O}(e) \\
& = \frac{p\, H_{\rm eff}^{(0)} H_{\rm EOB}^{(0)}}{A^2 \bar{D}\, p_r^{(1)}}+\mathcal{O}\left(e\right) .
\end{split}
\end{align}
With this, we see that even though $\dot{r}$ and $\partial r / \partial \zeta$ both vanish in the quasicircular limit, the rate of change of $t$ with respect to $\chi$ is finite and non-zero, and $\chi$ becomes the phase of the radial oscillation about the circular orbit. One can then divide the rates of change with respect to $t$ above by $d\chi/dt = 1/f_t^{(0)}$ to find the circular orbit limits of the right-hand sides of the equations of motion \eqref{eq:OOE_EoM_QC_Explicit} with respect to $\chi$.

\bibliography{EOBNITsQuasiCircularReferences}

\end{document}